\documentclass[%
 reprint,
superscriptaddress,
 amsmath,amssymb,longbibliography,
 aps
]{revtex4-2}

\usepackage{graphicx}
\usepackage{dcolumn}
\usepackage{bm}

\newcommand{\poincare}{Poincar\'{e} }

\begin{document}

\preprint{APS/123-QED}

\title{Waveguiding driven by the Pancharatnam-Berry phase}

\author{Chandroth P. Jisha}
 \email{jisha.chandroth.pannian@uni-jena.de}
\author{Stree Vithya Arumugam}%
\affiliation{%
   Friedrich Schiller University Jena, Institute of Applied Physics, Abbe Center of Photonics, Albert-Einstein-Str. 15, 07745 Jena, Germany
}%
\author{Lorenzo Marrucci}

\affiliation{
 Dipartimento di Fisica ``Ettore Pancini'', Universit\`{a} di Napoli Federico II, Complesso Universitario di Monte Sant’Angelo, Via Cintia, 80126 Napoli, Italy
}%

\author{Stefan Nolte}%
\affiliation{%
   Friedrich Schiller University Jena, Institute of Applied Physics, Abbe Center of Photonics, Albert-Einstein-Str. 15, 07745 Jena, Germany
}%
\affiliation{Fraunhofer Institute for Applied Optics and Precision Engineering IOF, Albert-Einstein-Str. 7, 07745 Jena, Germany}

\author{Alessandro Alberucci}%
\email{alessandro.alberucci@uni-jena.de}
\affiliation{%
   Friedrich Schiller University Jena, Institute of Applied Physics, Abbe Center of Photonics, Albert-Einstein-Str. 15, 07745 Jena, Germany
}%




\date{\today}

\begin{abstract}
We theoretically and numerically investigate the properties of waveguides based on the Pancharatnam-Berry phase, obtained by a longitudinally periodic rotation of the optic axis in a transversely-twisted birefringent medium. In this paper we study the case where the period of the longitudinal modulation is chosen so that a net accumulation of geometric phase in propagation occurs. First, the interplay between different contributions to the optical potential is addressed. Second, a continuous evolution of the polarization structure of the quasi-modes is observed in the numerical simulations. We explain it by a combination of plane-wave-based models and gauge transformations. We discover that, beyond the longitudinal oscillations, the polarization of the quasi-mode also varies through its cross-section. The analogies with respect to charged particles moving in a magnetic field are  outlined.

\end{abstract}

\maketitle


\section{\label{sec:intro}Pancharatnam-Berry phase in twisted anisotropic materials}

The propagation of plane waves in homogeneous anisotropic media is well understood: their refractive index depends on the direction of the electric field, with in general a non-parallel condition between the electric field $\bm E$ and the displacement vector $\bm D$ \cite{Yariv:1984}. Mathematically speaking, anisotropic materials are defined by a dielectric tensor $\bm \epsilon_D=\text{diag}\left(\epsilon_{x^\prime x^\prime},\epsilon_{y^\prime y^\prime},\epsilon_{z^\prime z^\prime}\right)$, where $x^\prime, y^\prime, z^\prime$ identify the principal dielectric axes. In uniaxial materials, the first two eigenvalues are identical and named $\epsilon_\bot$, whereas $\epsilon_{y^\prime y^\prime}=\epsilon_\|$ is the dielectric constant along the optic axis $\hat{n}=\hat{y}^\prime$. When the wavevector $\bm k$ is normal to the optic axis, the two independent eigenmodes are the extraordinary and the ordinary electric waves, perceiving respectively the refractive indices $n_{\|}=\sqrt{\epsilon_\|}$ and $n_\bot=\sqrt{\epsilon_\bot}$. This is the configuration used in waveplates, which  control the light polarization via the phase retardation $\Delta\phi= k_0\Delta n L$, where $\Delta n=n_\|-n_\bot$ is the birefringence, $k_0$ is the vacuum wavenumber, and $L$ is the length of the anisotropic material along the propagation direction $z$. From a mathematical point of view, the propagation of optical plane waves in anisotropic materials can be described using the Jones formalism, where a two-component vector fully determines the electromagnetic field \cite{Jones:1941}.\\
A surprising new effect arises when the Jones calculus  is applied to a twisted anisotropic material, i.e., a material whose optic axis varies on the transverse plane $xy$ orthogonal to the wavevector $\bm k$. If we name $\theta$ the angle between the optic axis and the axis $y$, when $\Delta \phi=\pi$ (half-wave plate, HWP) a circular polarized beam accumulates a transverse phase modulation given by $\pm 2\theta(x,y)$, the sign depending on the handedness of the impinging photons \cite{Bhandari:1997}. This additional phase term is a manifestation of geometric phase, an additional delay added to the dynamic phase (the optical path in optics) occurring when the Hamiltonian of a system is subject to a change in propagation \cite{Cohen:2019}. First introduced in a quantum mechanical framework and in the presence of a periodic evolution by Sir Michael Berry in 1984 \cite{Berry:1984}, a specific type of geometric phase was actually discovered by Pancharatnam 30 years earlier while studying polarized waves \cite{Pancharatnam:1956}. In the presence of a varying polarization along the propagation direction $z$, Pancharatnam found that an optical beam acquires a phase proportional to the corresponding area subtended by the polarization state trajectory on the \poincare sphere. When the polarization trace of a circularly polarized beam propagating in a wave plate is drawn on the \poincare sphere, it is evident that the phase term $\pm2\theta$ is a manifestation of the mechanism described by Pancharatnam. This phase is today called the Pancharatnam-Berry phase (PBP) in honor of its two fathers \cite{Jisha:2021}.\\ 
Probably due to technological constraints in manufacturing twisted anisotropic materials, the idea of wavefront manipulation through the PBP has not been pursued until the early 2000, the year in which the first experimental demonstration has been accomplished using sub-wavelength metallic gratings with a point-dependent orientation \cite{Bomzon:2001_2}. The idea has been applied some years later in liquid crystals, where the local optic axis can be controlled by a proper shaping of the boundary conditions \cite{Marrucci:2006,Marrucci:2006_1,Kim:2015}. The field literally exploded when wavefront shaping was demonstrated in metasurfaces, ultra-thin metamaterials featuring sub-wavelength structures. To observe PBP modulation, the basic elements of metasurface must lack rotational symmetry, thus mimicking the response of an anisotropic material \cite{Yu:2014,Arbabi:2015,Tymchenko:2015,Genevet:2017}. Currently, PBP is a central topic in modern optics, setting a new frontier for the control of light propagation \cite{Jisha:2021}. \\
As stated above, the phase modulation proportional to the local twist angle appears when the material is an infinitely thin HWP, that is, the propagation distance is negligible with respect to the Rayleigh distance of the beam. The interplay between diffraction and PBP has been investigated both in longitudinally invariant and periodically modulated twisted geometries, where the PBP action is modelled by effective potential(s) dependent on the local rotation angle $\theta$ \cite{Calvo:2007,Karimi:2009,Slussarenko:2016,Alberucci:2016}. In both cases, it has been demonstrated how the effective potential acting on the photons can be tailored to realize refractive index gradient-free optical waveguides \cite{Slussarenko:2016,Alberucci:2016}, with potential applications in topological photonics \cite{Abbaszadeh:2021}. In this paper we will use the more compact name Berry waveguide. Finally, the existence of the potential has been demonstrated experimentally in the nonlinear regime in liquid crystals \cite{Jisha:2019_1}. \\
Here we investigate theoretically and numerically the optical propagation in a twisted material, periodically modulated with a period $\Lambda=\lambda/\Delta n$ to allow the accumulation of PBP in propagation, with an approach that reminds of quasi-phase matching in nonlinear optics. We will discuss how the polarization structure of the localized quasi-mode evolves as the twisting of the material is increased. We will show how higher order effects, related with the non-adiabatic changes in the material parameters and mainly modelled via local gauge transformations, deeply impact light propagation. We will emphasize how the point-dependent twisting of the material is responsible for a very strong spin-orbit interaction, the latter being tunable with the maximum rotation angle applied to the medium. 


\section{\label{sec:model}Optical propagation in a periodic anisotropic structure}

Neglecting the longitudinal component along the propagation distance $z$, 
the electric field can be depicted as a two-component vector $\bm \psi=\left(E_x;\ E_y \right)$.  The approximate field $\bm \psi$ then obeys the vectorial Helmholtz equation $\nabla^2\bm \psi + k_0^2 \bm \epsilon \cdot \bm \psi=0$, where $\bm \epsilon$ is determined by the local twist angle
$\theta(x,y,z)$. Given the longitudinal component is neglected, hereafter we will restrict the dielectric tensor $\bm \epsilon$ to the transverse $xy$ components. Specifically, it is $\bm \epsilon= \bm R^{-1}(\theta)\cdot \bm \epsilon_D \cdot \bm R(\theta)$,  where $\bm{R}(\theta) = \left (   \cos\theta, \sin\theta;  -\sin\theta , \cos\theta  \right)$. The dielectric permittivity is given by 
$\epsilon_{ij}=\delta_{ij}\epsilon_\bot+\epsilon_a n_i n_j\ (i,j=x,y,z)$ \cite{Simoni:1997}, where $\bm \hat{n}$ is the unit vector along the local optic axis and $\epsilon_a=\epsilon_\|-\epsilon_\bot$ is the optical anisotropy.  Given we allow only for rotations of the optic axis in the plane $xy$, the relative dielectric permittivity tensor is
\begin{equation}
    \bm{\epsilon}= \epsilon_\bot \bm I + \epsilon_a \left(\begin{array}{cc}
      \sin^2 \theta & \cos\theta \sin\theta \\
      \cos\theta\sin\theta & \cos^2 \theta
    \end{array}\right).
\end{equation}  
Incidentally, in terms of the Pauli matrices $\bm \sigma_i\ (i=1,2,3)$ 
it is $\bm{R}(\theta)=e^{i\bm \sigma_2 \theta(x,y,z)} $, where we recall that $\sigma_2 =(0,-i;i,0)$. The two-component electric field obeys
\begin{equation}  \label{eq:spinorial_E}
 \nabla^2 \bm \psi + k_0^2 \left\{\overline{\epsilon} \bm I + \frac{\epsilon_a}{2} \left[ \bm \sigma_1 \sin(2\theta) - \bm \sigma_3 \cos(2\theta)\right] \right\}\cdot \bm \psi = 0,   
\end{equation}
where $\overline{\epsilon}= \left(\epsilon_\bot + \epsilon_\| \right)/2$. 
Terms proportional to the optical anisotropy $\epsilon_a$ can be rearranged in the form of a magnetic interaction
\begin{equation}
  H_{per}= \frac{\epsilon_a}{2} \left[ \bm \sigma_1 \sin(2\theta) - \bm \sigma_3 \cos(2\theta)\right]   = \frac{1}{2}   \bm{\sigma} \cdot \bm B_\mathrm{eff}, \label{eq:magnetic_field}
\end{equation}
where $\bm B_\mathrm{eff}=\epsilon_a\left[ \sin(2\theta) \hat{e}_1 - \cos(2\theta) \hat{e}_3\right]$ represents an effective magnetic field \cite{Fang:2012,Rechtsman:2013_1,Schine:2016}, here defined within a three-dimensional vector space spanned by unit vectors $\hat{e}_i\ (i=1,2,3)$. Following our definition, $\bm B_\mathrm{eff}$ is anti-parallel to $\hat{e}_3$ for $\theta=0$. Invariance to global rotation is automatically satisfied by the scalar product in Eq.~\eqref{eq:magnetic_field}. Finally, given that rotations of $180^\circ$ do not vary the optical properties of the anisotropic slab, the angle formed by $\bm B_\mathrm{eff}$ in the plane $e_1e_3$ is double the physical angle made by the optic axis on the transverse plane $xy$.  

Once rewritten in the paraxial limit, Eq.~\eqref{eq:spinorial_E} closely reminds the Pauli equation for a massive particle subject to a homogeneous scalar potential (term proportional to $\overline{\epsilon}$) and to a fictitious magnetic field of constant amplitude $\epsilon_a$, but changing its direction while lying on the plane $e_1e_3$. In optical terms, this shows that there are no refractive index gradients in this configuration. The effective magnetic field also explains the fundamental role played by geometric phase in driving the optical propagation \cite{Lin:2014_1,Jisha:2021}.\\
For the sake of simplicity, hereafter we focus on the (1+1)D case setting $\partial_y=0$.
To correctly apply the paraxial conditions, we rewrite the field $\bm \psi$ in an inhomogeneously rotated system  $ \bm \psi^\prime =\bm R [\theta(x,z)] \cdot \bm \psi$, i.e., we apply a local gauge transformation. A similar approach is used when describing the Majorana spin flip occurring for example in magnetic traps \cite{Sukumar:1997}.  Given the dielectric tensor is now diagonal everywhere, the light wave fulfills the following vectorial equation \cite{Slussarenko:2016}  
\begin{multline}
   \frac{\partial^2 \bm{\psi^\prime}}{\partial z^2} - i  \bm{\sigma_2} \cdot \left(2\frac{\partial\theta}{\partial z} \frac{\partial \bm{\psi^\prime}}{\partial z} + \frac{\partial^2\theta}{\partial z^2} \bm{\psi^\prime} \right)  + k_0^2 \bm{\epsilon}_D \bm{\psi^\prime} - \left(\frac{\partial\theta}{\partial z}\right)^2  \bm{\psi^\prime}    = \\
 -   \frac{\partial^2 \bm{\psi^\prime}}{\partial x^2} + \left(\frac{\partial\theta}{\partial x} \right)^2  \bm{\psi^\prime}   + i \frac{\partial^2 \theta}{\partial x^2}  \bm{\sigma_2}\cdot \bm{\psi^\prime} + 2i  \frac{\partial\theta}{\partial x} \bm{\sigma_2} \cdot \frac{\partial \bm{\psi^\prime}}{\partial  x}.  \label{eq:maxwell_rotated_inho}
\end{multline}
The left hand side (LHS) of Eq.~\eqref{eq:maxwell_rotated_inho} models the propagation of plane waves in a longitudinally-rotated twisted material, with no gradients along the transverse direction $x$. Let us now define the $2\times 2$ matrix $\bm N = \left(n_\bot,0;0,n_\| \right)$. The paraxial approximation (i.e., setting $\partial_z^2 \bm{\psi^\prime} =0$) is correctly applied to Eq.~\eqref{eq:maxwell_rotated_inho} if the transformation $\bm \psi^\prime = e^{ik_0\bm N z} \cdot \bm u$ is carried out, where $\bm u$ is the slowly varying vectorial envelope. Remarkably, the rotating field transformation factors out the different phase velocities of the ordinary and extraordinary components: for example, a field $\bm u$ featuring a circular polarization will conserve its polarization in propagation.\\
We now specialize our treatment to periodic modulations of the twisting angle along the propagation direction by setting $\theta(x,z)=H(z)\Gamma(x)$, with $H(z)=H(z+\Lambda)$. We further assume $\Lambda=\lambda/\Delta n$, where $\lambda/\Delta n$ is the birefringence period providing the natural oscillation of the optical polarization in the material (i.e., the full wave plate length). The equivalence between the natural oscillation and the external modulation allows a net accumulation of PBP in propagation \cite{Slussarenko:2016}.\\
As described by Eq.~\eqref{eq:maxwell_rotated_inho}, in first approximation optical propagation in twisted anisotropic materials has strong similarities with the same process in inhomogeneous isotropic materials: i) a diffraction operator tending to broaden the beam along the transverse direction; ii) a wavefront modulation proportional to the transverse gradient in the optical properties of the material. In our case the gradient is imposed on the twisting angle, and provides a point-dependent phase modulation associated with a change in the polarization with $z$. Actually, an intuitive model can be formulated by investigating the propagation of plane waves (i.e., in the absence of diffraction) in  materials that are treated as homogeneous along the transverse direction $x$ but change periodically along the propagation direction $z$. Physically speaking, this approach is exact for very slowly rotations of the optical axis along the transverse direction $x$. In the limit of small birefringence $\Delta n$, this case can be solved by applying the Jones' formalism to a stack of infinitely thin layers (see Appendix~\ref{sec:Jones_formalism}). In this limit the optical propagation depends on the phase retardation $\zeta=k_0 \Delta n z$. The interplay with diffraction can then be accounted for in a second stage.

\subsection{Plane-wave solution when the longitudinal modulation is a square wave}

Let us start from a brief summary of the circularly polarized (CP) plane wave propagation in an anisotropic material where the longitudinal modulation follows a square-wave function of duty cycle $50\%$. At the end of the first half period ($z=\Lambda/2$, HWP distance), the beam inverts its spin and acquires a PBP proportional to $2\Gamma(x)$. In $z=\Lambda/2$, the optic axis is flipped with respect to the $y-$axis (i.e., $\Gamma\rightarrow -\Gamma$), hence permitting the accumulation of an additional phase $2\Gamma(x)$ and, at the same time, closing the loop by returning to the original polarization. Thus, after propagating across a length $\Lambda$ the field returns to its initial polarization state, but has acquired a phase delay of geometric origin equal to $4\Gamma(x)$ \cite{Jisha:2021}. This cycle can then be repeated, leading to a progressive accumulation of this phase delay. The full behavior of the Stokes parameter is plotted in Fig.~\ref{fig:stokes_after_rotation} in Appendix~\ref{sec:plane_wave_model}. 

\subsection{Quasi-modes in the transversely-homogeneous case}
\label{sec:quasimodes_1D}
\begin{figure}
    \centering
    \includegraphics[width=\linewidth]{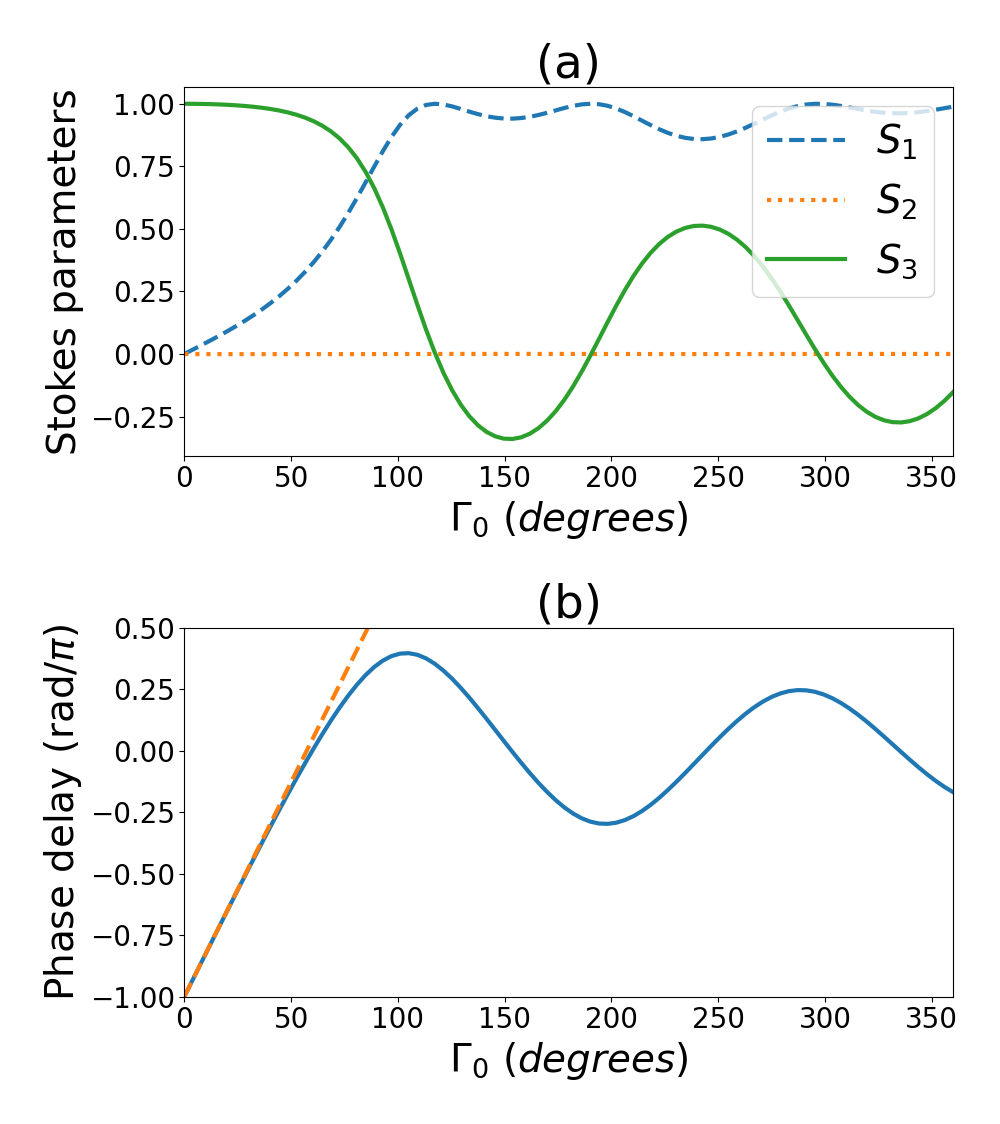}
    \caption{Properties of quasi-modes in a sinusoidally rotated anisotropic material. (a) Stokes parameters of the eigenmode (corresponding to the polarization assumed at the start of each longitudinal period) and (b) the corresponding geometric phase delay $\phi$ gained across a single rotation period versus the maximum rotation angle $\Gamma_0$. The dashed line in (b) corresponds to the geometric phase computed under the small rotation approximation, providing $\phi=\pi \Gamma_0$. Due to the symmetry of the system, another set of eigenmodes with opposite Stokes parameters and opposite phase delay exists.}
    \label{fig:plane_waves_eigenvectors}
\end{figure}

Due to the periodic nature of the system, eigenwaves of the system can be found considering one single oscillation period, $z\in [z_0, z_0+\Lambda]$. Hereafter we consider only sinusoidal waveforms for $H(z)$; we also fix $\Gamma(x)=\Gamma_0 f(x)$, where the peak of $f(x)$  is equal to unity. The numerically-computed eigenvectors and eigenvalues are plotted as a function of $\Gamma_0$ in Figure~\ref{fig:plane_waves_eigenvectors} (see Appendix~\ref{sec:Jones_formalism} for the employed numerical method). For a vanishing $\Gamma_0$, the polarization states move along the meridian of the \poincare sphere containing both the poles (CPs) and the diagonal/anti-diagonal linear polarization (defined with respect to the reference system $xy$). Stated otherwise, there is a sinusoidal oscillation of the Stokes parameters $S_2$ and $S_3$, while $S_1$ is null in every point of the path (see e.g. Fig.~\ref{fig:stokes_after_rotation} in Appendix~\ref{sec:plane_wave_model}). As $\Gamma_0$ assumes small but finite values, the trajectory moves away from the meridian and acquires a small component along $S_1$, see Fig.~\ref{fig:plane_waves_eigenvectors}(a) \cite{Hunter:2007}. Up to $\Gamma_0\approx45^\circ$, the $S_1$-value increase of  the eigenstate is linear with $\Gamma_0$. The growth of $S_1$ then gets steeper, with an inflection point around $\Gamma_0\approx 110^\circ$ and eventually reaching a local maximum around $\Gamma_0\approx 117^\circ$. After the local maximum, $S_1$ oscillates close to the maximum possible value $S_1=1$ in a quasi-periodic fashion. The oscillation period varies between $70^\circ$ and $90^\circ$. Remarkably, $S_2$ is always vanishing, no matter how large the rotation angle is. The accumulated phase delay plotted in Fig.~\ref{fig:plane_waves_eigenvectors}(b) is linearly increasing versus $\Gamma_0$ for small $S_1$, then undergoing an upwards bending around $S_1=0.25$. For large enough $S_1$, the accumulated PBP varies in a non-monotonic fashion, following the changes in the polarization, as first pointed out by Pancharatnam. \\
\begin{figure}
    \centering
    \includegraphics[width=\linewidth]{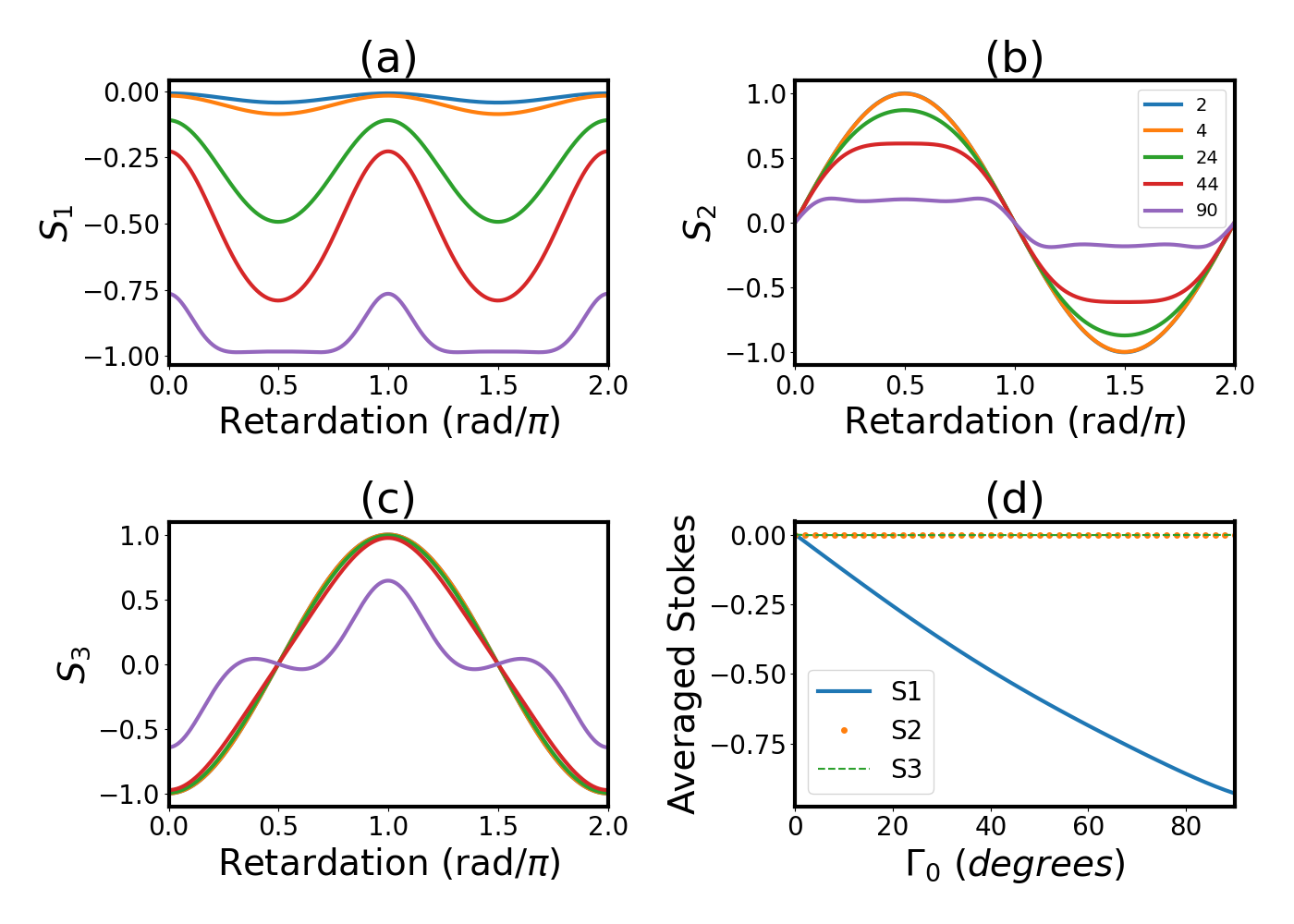}
    \caption{Evolution of the Stokes parameters (a) $S_1$, (b) $S_2$ and (c) $S_3$ versus the phase retardation $k_0\Delta n z$. The corresponding maximum rotation angles $\Gamma_0$ are reported in the legend in panel (b). (d) Average of the Stokes parameters over one birefringence length versus the angle $\Gamma_0$.}
    \label{fig:plane_waves_eigenvectors_evolution}
\end{figure}
In Appendix~\ref{sec:plane_wave_model} the quasi-modes of the Floquet-like system \cite{Shirley:1965} are investigated in the limit of small anisotropy by expanding the solution as a Bloch wave, $\bm  u(\zeta)= e^{i\beta \zeta} \sum_m {\bm u_m e}^{i m\zeta}$. The corresponding eigenvalue problem reads
\begin{multline}
\beta \bm u_m = -m \bm u_m +
    \frac{\Gamma_0}{4}[ 2\bm \sigma_2  \bm u_m + \left(\bm \sigma_2 - i\bm \sigma_1 \right) \bm u_{m-2} \\ + \left(\bm \sigma_2 + i\bm \sigma_1 \right) \bm u_{m+2}  ]. \label{eq:eigenproblem_bloch_main}
\end{multline}
According to Eq.~\eqref{eq:eigenproblem_bloch_main}, for small $\Gamma_0$ the quasi-modes are circularly polarized with a phase $\phi=2\pi \beta=\pm\pi\Gamma_0$ [Eq.~\eqref{eq:eigenproblem_bloch_main} provides $\beta=\pm \Gamma_0/2$], the sign being determined by the handedness of the CP wave (i.e., the photon spin). This is in agreement with the numerical results plotted in Fig.~\ref{fig:plane_waves_eigenvectors}(b) for $\Gamma_0$ up to $50^\circ$. The polarization of $\bm u$ is constant in propagation only in the rotated framework: when the transformation back to the laboratory framework is carried out, the CP will be retained only at the beginning and at the end of a birefringence length, whereas the Stokes vector will evolve periodically. As shown in Fig.~\ref{fig:stokes_after_rotation} in Appendix~\ref{sec:plane_wave_model}, $S_3$ versus $\zeta$ remains sinusoidal in this limit, whereas $S_2$ follows  sinusoidal curves which are flattened around $z=\Lambda/4$ and $z=(3/4)\Lambda$, with a corresponding increase in $|S_1|$ in the same regions. This is confirmed for $\Gamma_0$ up to $50^\circ$ by the exact evolution along $z$ of the polarization plotted in Fig.~\ref{fig:plane_waves_eigenvectors_evolution}. An additional effect observed in the numerical solution is that the value of $S_1$ in  $z=0$ is not vanishing, see Fig.~\ref{fig:plane_waves_eigenvectors}(a) and Fig.~\ref{fig:plane_waves_eigenvectors_evolution}(a). This can be explained from Eq.~\eqref{eq:eigenproblem_bloch_main} once the terms $u_{\pm 1}$ are accounted for, see Appendix~\ref{sec:plane_wave_model}.
Even in this limit, the associated eigenvalue $\beta$ (i.e., the local optical delay $\phi$) remains unperturbed, in accordance with the full simulations for $\Gamma_0<50^\circ$. The higher-order harmonics $u_m$ ($|m| > 1$) become relevant when $\Gamma_0>50^\circ$, as witnessed by a strong deformation in $S_3$ versus $z$, see Fig.~\ref{fig:plane_waves_eigenvectors_evolution}(c). Finally, Fig.~\ref{fig:plane_waves_eigenvectors_evolution}(d) shows how only the average value of $S_1$ is different from zero, whereas $S_2$ and $S_3$ conserve a periodic motion with a vanishing average. 


\subsection{Coupling with diffraction}

The terms on the RHS of Eq.~\eqref{eq:maxwell_rotated_inho} stem from the Laplacian operator, i.e., they originate from the natural spreading of light in space. In the case of twisted anisotropic materials, complicated effects arise from the coupling between neighbouring points in the transverse plane. Indeed, a wave of a given linear polarization can solely correspond to a local eigensolution (extraordinary or ordinary polarized) of Maxwell's equations. Diffraction transports a portion of this local eigensolution to adjacent regions where the optic axis is differently oriented, in turn leading to a continuous local change in the beam polarization and phase. In agreement with the case of plane waves discussed in the previous section, a localized solution of the electromagnetic equation in this geometry needs to be periodic along $z$. The purpose of the current subsection is to find a simplified equation for the continuous component of the optical field using the normalized coordinates $\zeta=k_0\Delta n z$ and $\eta=x/\lambda$. After making the further gauge transformation 
$\bm u=e^{i\bm \sigma_1 \Gamma/2}\cdot \bm v $, in the limit $\gamma=\Delta n/\overline{n}\ll 1$ the continuous wave (CW) component of the field $\bm{v}$ satisfies the following Pauli-like equation (see Appendix~\ref{sec:transverse_coupling})
\begin{multline}
   i \gamma  \frac{\partial \bm{v}_0}{\partial \zeta} = 
    -\frac{1}{8\pi^2  \overline{n}^2} \frac{\partial^2 \bm v_0}{\partial \eta^2}
   -\frac{\gamma \Gamma}{2}
      \left[ \cos\left(\Gamma \right)  \bm \sigma_2 + \sin \left(\Gamma \right)  \bm \sigma_3  \right] \cdot \bm v_0  \\ + \frac{1}{32\pi^2  \overline{n}^2} 
      \left( \frac{\partial \Gamma}{\partial \eta}\right)^2 \bm v_0 .
  \label{eq:Pauli_model}
\end{multline}
In agreement with the plane-wave model \cite{Slussarenko:2016}, a spin-dependent phase modulation proportional to the local amplitude of the rotation angle $\Gamma(x)$ is acting on the beam. The gauge transformation modifies the spin-orbit coupling due to the multiplication between the original operator $\bm \sigma_2$ and the gauge operator $e^{i\bm \sigma_1 \Gamma(x)/2}$, in turn introducing a term containing $\bm \sigma_3$ and proportional to $\sin(\Gamma)$. The Stokes vector of the resulting structured beams in the rotated framework shows a non-vanishing component $S_1$, beyond the dominant circular polarization component given by $S_3$. The ratio between the two components of the Stokes vector also varies along the beam cross-section. \\
To conclude this section, we discuss the effect of the gauge transformation on the real polarization of the quasi-modes. The CW component in the rotated framework reads
\begin{equation}
    \bm u_0 = \cos\left(\frac{\Gamma}{2} \right) \bm v_0 +i\sin\left(\frac{\Gamma}{2}\right) \bm\sigma_1 \cdot \bm v_0 . \label{eq:quasimode_after_gauge}
\end{equation}
According to Eq.~\eqref{eq:quasimode_after_gauge}, the polarization of a portion of the quasi-mode [proportional to $\cos\left(\frac{\Gamma}{2} \right)$] found from Eq.~\eqref{eq:Pauli_model} remains unvaried after the gauge transformation. On the other side, the remaining part proportional to $\sin\left(\frac{\Gamma}{2} \right)$ is subject to a flip of its spin, i.e, the sign of the Stokes vector is inverted.
When $v\bm _0$ is CP, the phase difference between the two components of $\bm u_0$ is $\pm \Gamma+\pi/2$: as angle $\Gamma$ spans from zero to $\pi/2$, the beam polarization shifts from circular to linear diagonal or anti-diagonal polarization, assuming an elliptical polarization during the transition. For small $\Gamma$, Eq.~\eqref{eq:quasimode_after_gauge} becomes $\bm u_0\approx \bm v_0 - 0.5 \Gamma(x) \bm \sigma_1 \cdot \bm v_0$.

\section{\label{sec:simulations} Numerical simulations}

We simulated the behavior of light in a twisted anisotropic material by using a combination of FDTD (Finite Difference Time Domain) and FEM (Finite Element Method) software. For FDTD, we used the open source code MEEP \cite{Oskooi:2010}. For FEM, we used the commercial software COMSOL Multiphysics\textsuperscript{\textregistered} \cite{comsol}. Both the numerical simulators solve the complete Maxwell equations, thus accounting for the full vectorial nature of the field and for light rays propagating at wide angles with respect to the main carrier propagating along $z$. Here in the main text we will present solely the results calculated with the FDTD code. Details of FDTD simulations are provided in Appendix~\ref{sec:appendix_FDTD}, whereas the comparison with FEM results is carried out in Appendix~\ref{sec:appendix_FEM}.

\subsection{Potential and quasi-modes}

\begin{figure}
    \centering
    \includegraphics[width=\linewidth]{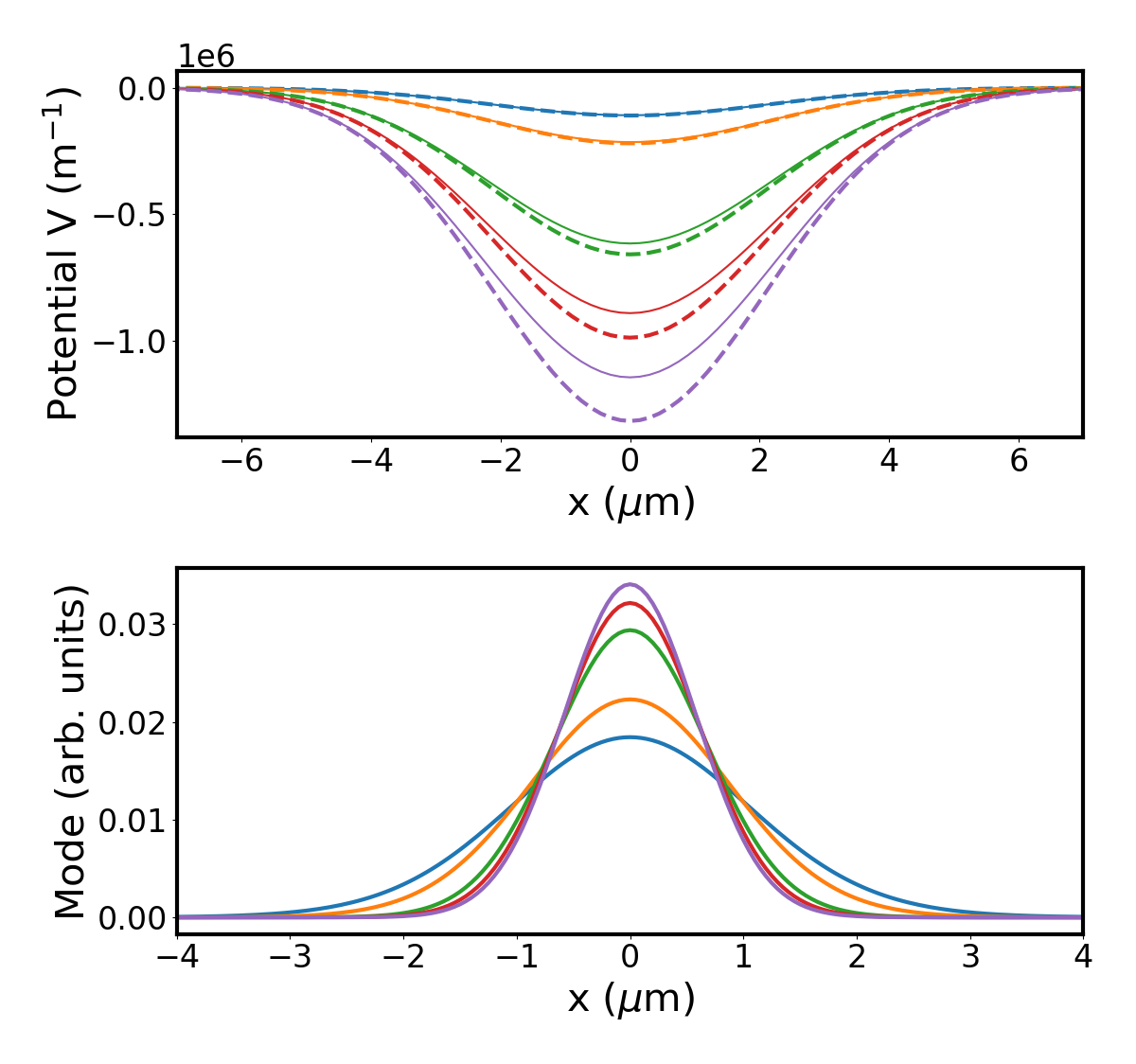}
    \caption{Photonic potential $V$  (a) and the corresponding fundamental mode (b, the intensity profile is shown) versus $x$. From shallower to deeper potential well (corresponding to a narrower fundamental mode), the maximum rotation angle $\Gamma_0$ is 10$^\circ$ (blue), 20$^\circ$ (orange), 60$^\circ$ (green), 90$^\circ$ (red), and 120$^\circ$ (magenta). In (a) solid and dashed lines correspond to the full potential evaluated from the entire Eq.~\eqref{eq:potential_first_order} or only its first term, respectively.}
    \label{fig:potential_modes}
\end{figure}

As input condition for the numerical simulations, we do not consider a generic Gaussian profile, but we instead prefer the quasi-mode profile predicted in Ref.~\cite{Slussarenko:2016} using a simplified theoretical model. This approach allows us to directly address the validity range of the two models (i.e., the model in Ref.~\cite{Slussarenko:2016} and the one discussed in this paper) in describing PBP-based optical waveguides. To first approximation the quasi-modes are CP modes subject to the following spin-dependent potential \cite{Slussarenko:2016}
\begin{equation} \label{eq:potential_first_order}
    V(x)= -\frac{S_3^{(0)} k_0 \Delta n }{2} \Gamma(x)+\frac{1}{4\overline{n}k_0} \left[ \left(\frac{\partial\Gamma}{\partial x} \right)^2 + k_0^2 \left( \Delta n\right)^2  \Gamma^2(x)  \right].
\end{equation}
The quantity $S_3^{(0)}$ is the third Stokes parameter sampled at the beginning of the longitudinal sinusoidal oscillation. A shift of $\pi$ in the sine (i.e., HWP longitudinal shift in the real space) yields a change in sign in the first term, i.e., the photon spin corresponding to waveguiding is switched.
Equation~\eqref{eq:potential_first_order} is the effective potential once the light propagation is recast for the scalar field $A$ in the form $i\partial_z A =-\left[ 1/\left(2\overline{n}k_0 \right)\right] \partial^2_x A + V A$. This means that light is attracted towards regions where $V$ is lower, in agreement with the quantum mechanical convention. 
The three terms composing the potential $V$ have a simple physical interpretation. The first term comes from the net accumulation of PBP due to the periodic longitudinal rotation of the optic axis. The second term and the third terms are Kapitza-like terms proportional to the square of the gradient of the rotation angle $\theta$ \cite{Alberucci:2016}. Essentially, a periodic modulation of the phase generates a local modulation of the transverse wavevector $k_x$, yielding a local modulation on the equivalent kinetic energy due to its dependence on the square of $k_x$. In agreement with Eq.~\eqref{eq:Pauli_model}, the term depending on the longitudinal derivative is $O\left[\left( \Delta n\right)^2\right]$, and can be neglected in the adiabatic limit. In practice, for a fixed material the approximation will start to fail for large enough twisting angle, given that this phase term depends quadratically on $\Gamma_0$. Hereafter we set the wavelength to $\lambda=1~\mu$m and the birefringence to $\Delta n=0.2$. The longitudinal shape of the modulation is kept sinusoidal in the remainder of the paper. The transverse distribution of the optic axis is assumed to be Gaussian by setting $\Gamma(x)=\Gamma_0 \exp{\left[-\left(x^2/w^2_D\right)\right]}$. Figure~\ref{fig:potential_modes} summarizes the behavior of the potential. The first term in Eq.~\eqref{eq:potential_first_order} is the most important term, assuming a confining or a repelling nature according to the sign of the impinging wave. The term proportional to $(\partial_x\Gamma)^2$ takes a typical W-shape and is negligible with respect to the other two terms for $w_D>1~\mu$m. Finally, the term proportional to $\Gamma^2$ is intrinsically defocusing (i.e., a positive hump), achieving an amplitude of about $20\%$ of the overall potential for $\Gamma_0=120^\circ$. This is visible in Fig.~\ref{fig:potential_modes}(a), where the potential versus $x$ for different $\Gamma_0$ is plotted. \\ 
To address the confinement strength of the photonic effective potential, the value of $V$ can be transformed into an effective gradient in the refractive index $\delta n$. Recalling that $V\approx - 2\overline{n}k_0 \delta n$, we obtain for example $\delta n \approx 0.05$ for $V=1\times 10^6$m$^{-1}$. In Fig.~\ref{fig:potential_modes}(b) the corresponding fundamental quasi-mode width versus $x$ for several $\Gamma_0$ is shown. In our case the beam width - defined as $w=2\sqrt{\int x^2 I(x)dx/\int{I(x)dx}}$- spans from $5.4~\mu$m at $\Gamma_0=1^\circ$, to $1.3\mu$m at $\Gamma_0=90^\circ$.

\begin{figure*}
    \centering
    \includegraphics[width=\linewidth]{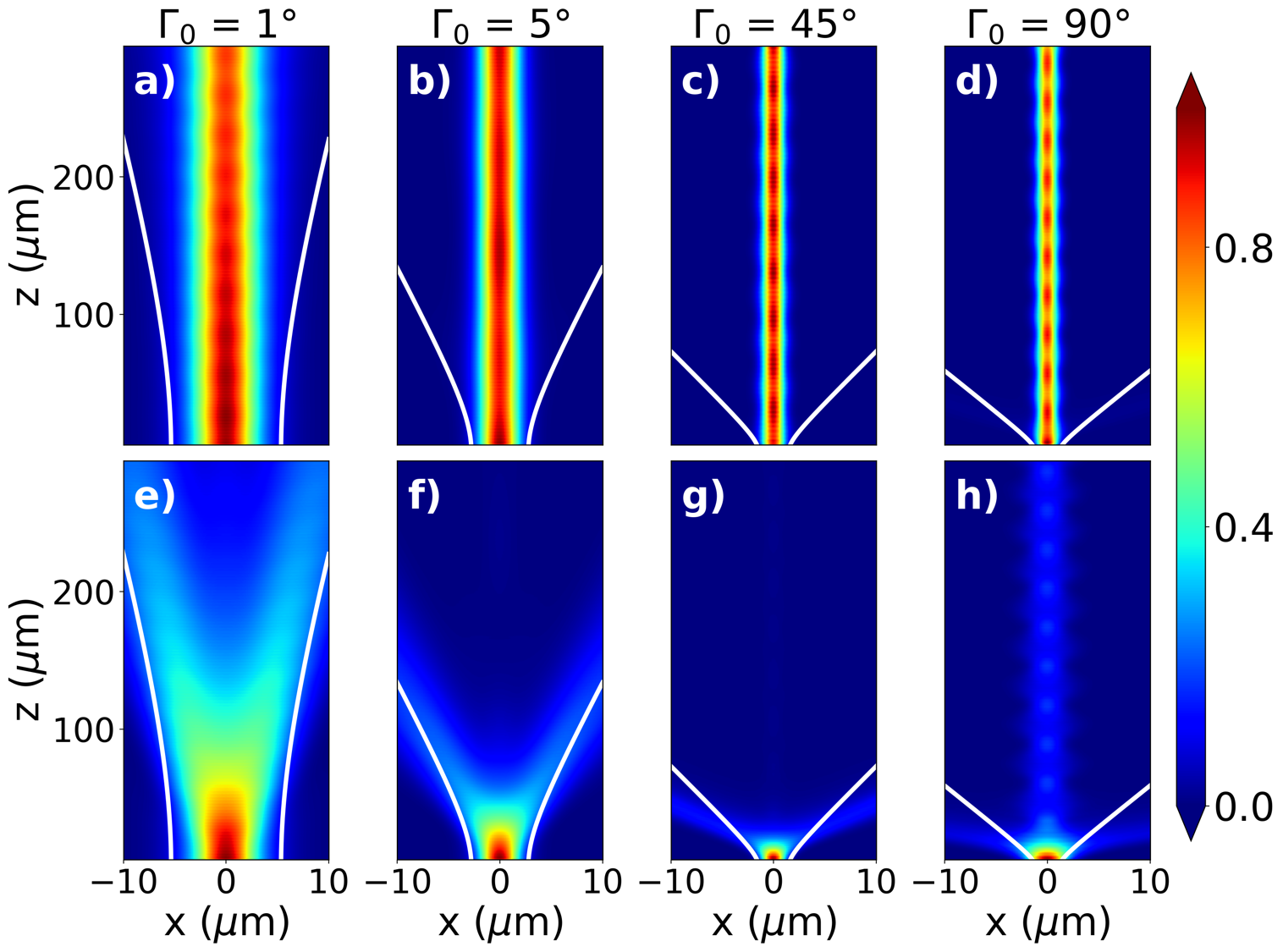}
    \caption{Time-averaged intensity distribution calculated via FDTD simulations for (a-d) RCP and (e-h) LCP input polarization for $w_D=3~\mu$m and increasing $\Gamma_0$ from left to right. The scalar eigenfunctions of the potential given by Eq.~\eqref{eq:potential_first_order} are used as the transverse shape of the input. The white solid lines represent the spreading that would occur in the case of a homogeneous material. Finally, the anisotropic material starts in $z=2~\mu$m.}
    \label{fig:FDTD_survey}
\end{figure*}


\subsection{Propagation of quasi-modes}
\begin{figure}
    \centering
    \includegraphics[width=\linewidth]{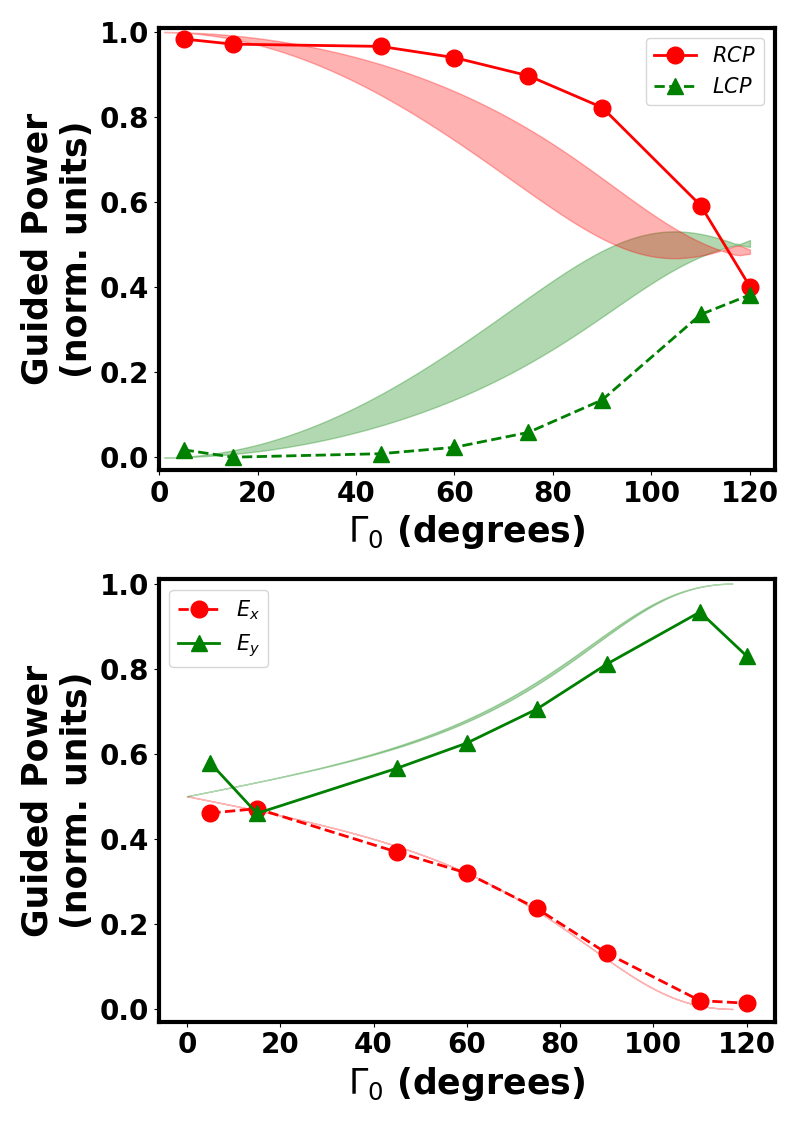}
    \caption{Power coupled to the Berry waveguide for input circular (a), and linear vertical and horizontal (b) polarizations versus the maximum rotation angle $\Gamma_0$. Symbols are values extrapolated from the FDTD simulations, whereas the shaded regions are the theoretical predictions from Eq.~\eqref{eq:quasimode_after_gauge}. Theoretical predictions corresponds to a surface because we are considering a range for the possible width of the quasi-mode, see the main text. The waveguide parameters are the same of Fig.~\ref{fig:FDTD_survey}. The guided power is measured in $z=200~\mu$m by integrating the intensity around the origin $x=0$ on a window of overall size $10~\mu$m. }
    \label{fig:coupling}
\end{figure}

At the entrance of the twisted material we used the quasi-mode calculated from Eq.~\eqref{eq:potential_first_order} (see Appendix~\ref{sec:appendix_FDTD} for the employed procedure). A survey of the intensity distribution versus the twisting angle $\Gamma_0$ is provided in Fig.~\ref{fig:FDTD_survey}.  In agreement with the strong spin-orbit coupling of our system, the general behavior for $\Gamma_0$ lower than $90^\circ$ strongly depends on the wave handedness: RCP (right CP) undergoes a net confinement while propagating [Fig.~\ref{fig:FDTD_survey}(a-d) shows the case $w_D=3~\mu$m; additional simulations not shown here demonstrate that an analogous behavior is found for larger $w_D$], whereas the LCP (left CP) waves spread more than would occur in a homogeneous cell, see Fig.~\ref{fig:FDTD_survey}(e-h). In each panel the white solid lines show the corresponding unconstrained diffraction (width $1/e^2$). With reference to the confined case, the quasi-modes are a very good approximation: the envelope of the beam propagates with very small oscillations for $\Gamma_0$ up to $90^\circ$. At these large angles, the trapping is retained, but the observed breathing amplitude is quite large. Indeed, at large $\Gamma_0$ a new propagation regime arises: the optical propagation does not significantly depend anymore on the input helicity. A precursor of this behavior is already visible in Fig.~\ref{fig:FDTD_survey}(h), where at $\Gamma_0=90^\circ$ an appreciable  portion of the input power is guided, even for the polarization where defocusing takes place for lower angles. This behavior is in remarkable agreement with the plane wave model plotted in Fig.~\ref{fig:plane_waves_eigenvectors}, where the accumulation of PBP stops to monotonically increase for $\Gamma_0>110^\circ$. The dependence of the power coupled to the quasi-mode versus $\Gamma_0$ and the input polarization is plotted in Fig.~\ref{fig:coupling}. For small angles the whole system response is analogous to a circular birefringent material [see Fig.~\ref{fig:coupling}(a)], where the confinement/defocusing of the beam depends on the handedness at the input. This is similar to what happens in cholesteric liquid crystals, where a helically-twisted uniaxial behaves at large scales like a circularly birefringent material \cite{Simoni:1997}. The two curves for different spins starts to flex towards each other around  $\Gamma_0\approx 50^\circ$, eventually crossing in $\Gamma_0=120^\circ$. Figure~\ref{fig:coupling}(b) shows the guided power when the input polarization is linearly polarized. At small rotations the behavior is almost polarization-independent, in agreement with Fig.~\ref{fig:coupling}(a). As the twisting gets larger the two curves diverge from each other in a symmetric way. Stated otherwise, the degeneracy between the vertical (parallel to $y$) and the horizontal (parallel to $x$) polarization is broken. Around $\Gamma_0=120^\circ$ the optimal coupling to the quasi-mode occurs for the vertical polarization, whereas the orthogonal polarization reaches its maximum broadening due to the presence of a repelling potential. The general trend of the FDTD simulations (lines with symbols in Fig.~\ref{fig:coupling}) is in qualitative agreement with Eq.~\eqref{eq:quasimode_after_gauge} (shaded regions in the same figure): the polarization-dependence of the guiding effect strongly depends on the twisting angle due to the local gauge transformation, or, in more physical terms, due to the strong transverse coupling between regions with different twisting, ultimately induced by the natural tendency of light to diffract.  More in detail, on the theoretical side we compute the overlap integral between a non-structured input beam and the quasi-mode in a simplified manner. We take a given Gaussian beam at the input, with an $x-$independent polarization, selected as indicated by the legends in Fig.~\ref{fig:coupling}. To find an approximation for the structured localized mode, the transformation given by Eq.~\eqref{eq:quasimode_after_gauge} is then applied to this beam (i.e., the latter is $\bm v_0$ in this case), but with a polarization given by the plane wave model computed for each value of $\Gamma_0$, see Fig.~\ref{fig:plane_waves_eigenvectors}. The final step is to compute the overlap between the two spinors. We stress that: i) to account for the variations in the width of the quasi-mode versus $\Gamma_0$ (see Fig.~\ref{fig:potential_modes}), we consider two different widths for the quasi-mode -$1~\mu$m and $5~\mu$m- corresponding to the edges of the shaded region; ii) the exact $\bm v_0$ is already a structured beam, whereas here its polarization is taken to be invariant through its cross-section.  

\begin{figure}
    \centering
    \includegraphics[width=\linewidth]{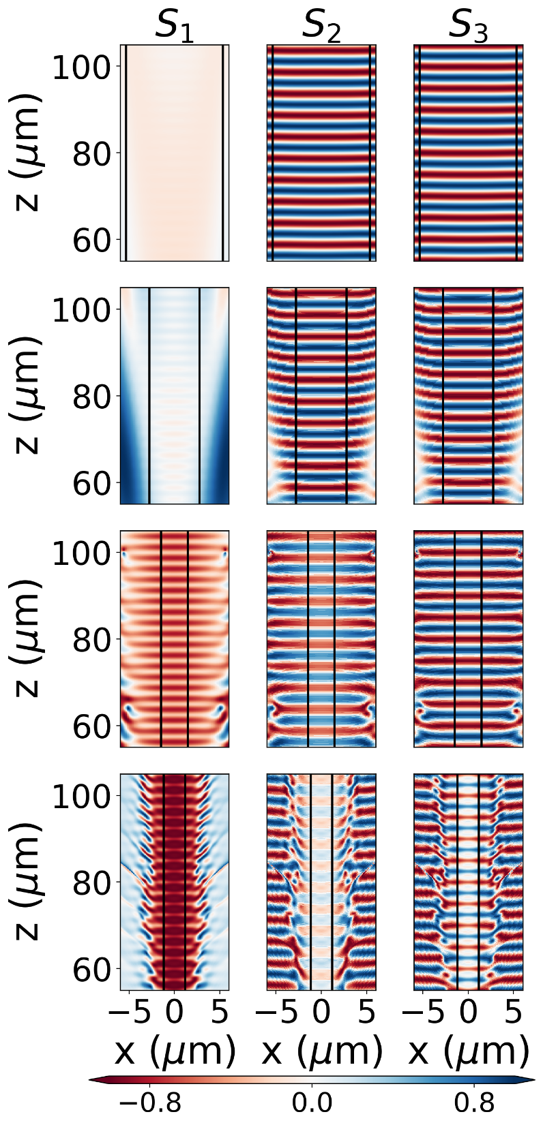}
    \caption{Distribution of the Stokes parameters on the portion of the plane $xz$ nearby the input interface extracted from FDTD simulations, plotted for $\Gamma_0=1^\circ$, $5^\circ$,  $45^\circ$ and $90^\circ$ from top to bottom, respectively. The input is a RCP mode with shape found from the potential Eq.~\eqref{eq:potential_first_order}.}
    \label{fig:map_Stokes_input}
\end{figure}

\subsection{Full characterization in terms of Stokes parameters}
\begin{figure}
    \centering
    \includegraphics[width=\linewidth]{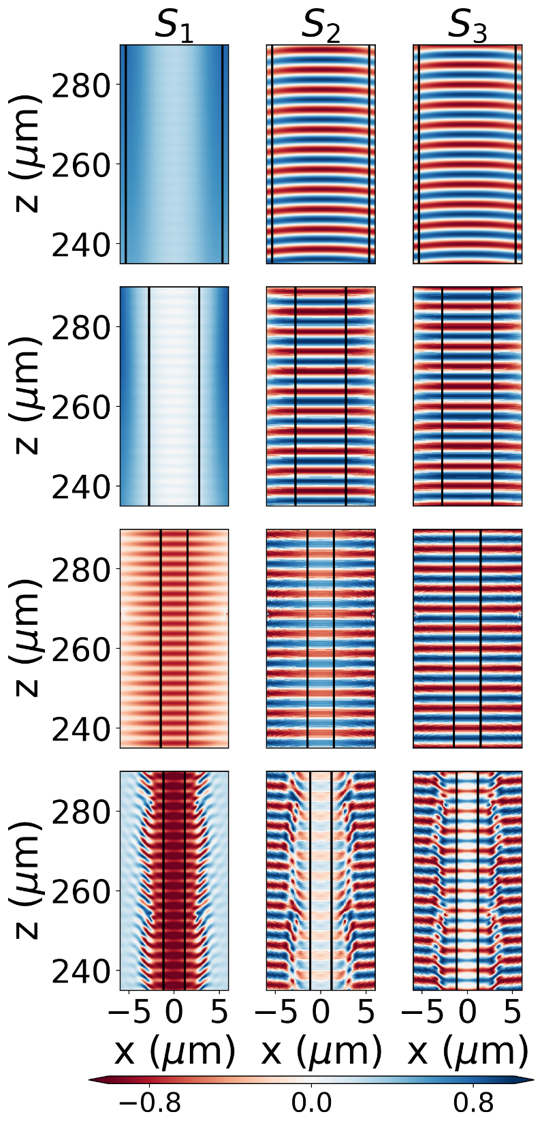}
    \caption{As in Fig.~\ref{fig:map_Stokes_input}, but at the end of the numerical grid, i.e., for $z>250~\mu$m.}
    \label{fig:map_Stokes_output}
\end{figure}

A deeper understanding on the physical mechanism behind the light confinement is achieved when the Stokes parameters of the propagating beams are plotted. Figure~\ref{fig:map_Stokes_input} and \ref{fig:map_Stokes_output} show the Stokes parameters corresponding to the trapped beam plotted in Fig.~\ref{fig:FDTD_survey}(a-d). The Stokes parameters are shown in proximity of the input interface (Fig.~\ref{fig:map_Stokes_input}) and deep inside the waveguide to show the effects of the mode coupling and the stationary localized wave, respectively.  
For very small angles ($\Gamma_0=1^\circ$), the mode computed from Eq.~\eqref{eq:potential_first_order} describes very well the propagating quasi-mode: the two Stokes parameters $S_2$ and $S_3$ vary sinusoidally with a period given by $\Lambda$ and a relative shift of a quarter of period, $\Lambda/4$, whereas $S_1$ is negligibly small. For $\Gamma_0=5^\circ$ the situation is very similar, except for the appearance of a non-vanishing $S_1$, in accordance with Fig.~\ref{fig:plane_waves_eigenvectors}(a). For $\Gamma_0=45^\circ$ a discrepancy in the polarization at the input interface is observed, with the emission of polarized radiation modes. In the bulk the sinusoidal variation of $S_2$ and $S_3$ is observed, but, unlike for smaller angles, $S_1$ is quite large, and encompasses a large $z-$invariant value superposed with a smaller sinusoidal oscillation of period $\Lambda$. For $\Gamma_0=90^\circ$ the coupling gets worse, with the periodicity being lost near the input interface. The oscillatory behavior of $S_2$ and $S_3$ is recovered into the bulk, although now the dominant component is $S_1$, the latter behaving similar to what is predicted by the plane wave model plotted in Fig.~\ref{fig:plane_waves_eigenvectors_evolution}(a). A large (about $180^\circ$) phase shift of the longitudinal oscillation between the center and the tails of the guided mode is observed for all the three Stokes parameters, even when the stationary regime is achieved: the quasi-mode is thus structured even along the transverse direction. The described dynamics confirms that the polarization of the quasi-mode follows at least qualitatively Eq.~\eqref{eq:quasimode_after_gauge}, and that the plane wave approach to calculate the phase delay shown in Fig.~\ref{fig:plane_waves_eigenvectors}(b) is quite reliable even in the presence of a local twisting. We thus evince that the breathing behavior observed in the intensity profile (Fig.~\ref{fig:FDTD_survey}) is due to a mismatch between the approximated quasi-mode (pseudo-scalar) and the real mode, the latter being highly structured both along the longitudinal and the transverse direction.

\section{Summary of the main results}

This work contains the following main results:

\begin{enumerate}
    \item In a transversely invariant but longitudinally rotated anisotropic material, it is possible to control all the Stokes parameters using a HWP-long sample, see Fig.~\ref{fig:plane_waves_eigenvectors} and Fig.~\ref{fig:plane_waves_eigenvectors_evolution}. As a direct consequence, the quasi-mode of a Berry waveguide is not purely circularly polarized. In particular, a constant component $S_1$ appears as the rotation is increased, a fact confirmed by the full numerical simulations of the Maxwell's equations.
    \item The transverse coupling due to diffraction in a transversely-inhomogeneous twisted sample can be modelled using  point-dependent gauge transformations. The transformation then yields the appearance of a Kapitza potential proportional to the transverse gradient of the twist, and of a point-dependent rotation of the polarization, see Eq.~\eqref{eq:quasimode_after_gauge}. This is another factor making the quasi-mode a fully structured beam \cite{Forbes:2021}, even in the transverse plane. Accordingly, in the numerical simulations the polarization of the localized wave is transversely variant, and the coupling between circularly polarized inputs and the quasi-modes drastically changes as the twisting ramps up.
    \item Despite the changes in the polarization described in the two previous points, the mode profile described by solving the scalar equation with the potential given by Eq.~\eqref{eq:potential_first_order} provides a very good approximation for the fundamental quasi-mode. First, the transverse Kapitza potential is negligible if sub-wavelength twisting of the material is left out. From Fig.~\ref{fig:potential_modes}, the interplay between the accumulated PBP and the longitudinal Kapitza effect determines the light propagation. Given that the Kapitza term is quadratic in the rotation angle, the PBP dominates at low angles, whereas the defocusing contribution of the Kapitza term becomes more and more relevant as the rotation is increased.
    \item For very large rotation angles, several new effects come into play. Even in the adiabatic limit (small anisotropy $\Delta n$), the accumulated PBP is no more monotonic given that the polarization path on the \poincare sphere becomes very complex and irregular. The real propagation is way more complex, as several new terms contribute as the variation speed of the polarization gets faster, see Appendix~\ref{sec:plane_wave_model}. To mention only a single effect, in the adiabatic limit the longitudinal Kapitza effect is absent, see Fig.~\ref{fig:plane_waves_eigenvectors} and Appendix~\ref{sec:plane_wave_model}. 
\end{enumerate}

\section{Conclusions}

In this paper we investigated theoretically and numerically the waveguiding observed in a periodically twisted anisotropic material and based upon a transverse gradient in the Pancharatnam-Berry phase. With respect to our previous work, we improved the theory by accounting for higher order effects, mainly including the fact that the guided modes feature a point-dependent polarization even across the transverse plane. 
For small angles, a purely circular polarized beam approximates well the confined mode, the transverse shape of the beam being in good agreement with the scalar potential originating from the PBP. For larger rotations, all the three Stokes parameters (including $S_1$) are not vanishing, in disagreement with the intuitive picture based upon a plane wave in the presence of a longitudinal modulation in the form of a square wave. Furthermore, as the rotation increases the helicity of the quasi-mode starts to flip and the mode to be strongly structured along its cross-section. \\ 
Although already observed in the nonlinear regime \cite{Jisha:2019_1}, the experimental realization of continuous PBP waveguides in the linear regime is the next step: different approaches to achieve this aim are currently pursued, including photo-polymerization of liquid crystals \cite{Kim:2015,Tartan:2017,He:2019}. multi-stack of inhomogeneously rotated liquid crystals plates \cite{Berteloot:2020}, and femtosecond writing of transparent materials \cite{Sakakura:2020}. As pinpointed in this article, these waveguides would support structured modes \cite{Forbes:2019}, thus representing an important advance in the current research about multi-modal optical communications \cite{Forbes:2021,Willner:2021}, both in the classical \cite{Milione:2015} and in the quantum regime \cite{Nagali:2009_1}.
In a broader physical perspective, our paper confirms a strict relation between twisted anisotropic media and propagation of charged particles in a magnetic field, proposing this optical platform as a promising candidate for the theoretical and experimental investigation of gauge-related and spin-orbit effects in an optical system \cite{Bliokh:2007,Alberucci:2010,Fang:2013,Liu:2015,Chen:2019_3,Lumer:2019,Brosco:2021,Huang:2022}.

\begin{acknowledgments}
C.P.J. has received funding from the European Union’s Framework Programme for Research and Innovation Horizon 2020 under the Marie Sklowdowska-Curie Grant Agreement No. 889525. S.V.A. is part of the Max Planck School of Photonics supported by BMBF, Max Planck Society, and Fraunhofer Society. This work is supported by the DFG Collaborative Research Center "NOA – Nonlinear Optics down to Atomic scales", Grant No. SFB 1375. The computational experiments were performed on resources of Friedrich
Schiller University Jena supported in part by DFG grants INST 275/334-1 FUGG and INST 275/363-1 FUGG. 
\end{acknowledgments}

\appendix

\section{Plane wave propagation in longitudinally twisted materials}
\label{sec:plane_wave_model}

We first apply the SVEA (Slowly Varying Envelope Approximation) to Eq.~\eqref{eq:maxwell_rotated_inho} by setting $\bm \psi^\prime = e^{ik_0\bm N z} \cdot \bm u$, where $\bm N = \left(n_\bot, 0;0,n_\| \right)$ \cite{Slussarenko:2016}. We obtain
\begin{multline}
   2ik_0 \bm N\cdot e^{ik_0\bm N z} \cdot \frac{\partial \bm{u}}{\partial z} \\- i  \bm{\sigma_2} \cdot \left(2 i k_0\frac{\partial\theta}{\partial z} \bm N \cdot e^{ik_0\bm N z} \cdot \bm u + \frac{\partial^2\theta}{\partial z^2} e^{ik_0\bm N z} \cdot \bm{u} \right) \\  - \left(\frac{\partial\theta}{\partial z}\right)^2 e^{ik_0\bm N z}  \cdot \bm{u}=0.    
  \label{eq:rotated_SVEA_PW}
\end{multline}
After multiplying both the sides of Eq.~\eqref{eq:rotated_SVEA_PW} by $e^{-ik_0\bm N z}$, we find that
\begin{multline}
   2ik_0 \bm N \cdot \frac{\partial \bm{u}}{\partial z} + 2  k_0\frac{\partial\theta}{\partial z} \bm \tilde{\sigma}_2(z) \cdot \bm N  \cdot \bm u - i \frac{\partial^2\theta}{\partial z^2}\bm \tilde{\sigma}_2(z) \cdot \bm{u}  \\  - \left(\frac{\partial\theta}{\partial z}\right)^2  \cdot \bm{u}=0,    
  \label{eq:rotated_SVEA_PW_2}
\end{multline}
where
\begin{equation}
    \bm \tilde{\sigma}_2(z) = \cos\left({k_0 \Delta n z} \right) \bm \sigma_2 + \sin\left({k_0 \Delta n z} \right) \bm \sigma_1. 
\end{equation}
$\bm \tilde{\sigma}_2$ oscillates along $z$ with the same period given by the material birefringence.
We want to express Eq.~\eqref{eq:rotated_SVEA_PW_2} solely in terms of Pauli matrices. At this purpose we set $\bm N=\overline{n} \bm I -\frac{\Delta n}{2} \bm \sigma_3$, where $\overline{n}=\left(n_\bot + n_\| \right)/2$ is the refractive index perceived by a circular polarization. Direct substitution into Eq.~\eqref{eq:rotated_SVEA_PW_2} yields
\begin{multline}
   2ik_0 \left(\overline{n} \bm I -\frac{\Delta n}{2} \bm \sigma_3 \right) \cdot \frac{\partial \bm{u}}{\partial z} + 2  k_0 \frac{\partial\theta}{\partial z} \left[F(z)\bm \sigma_2 + G(z) \bm \sigma_1 \right]  \cdot \bm u \\ - i \frac{\partial^2\theta}{\partial z^2}  \left[ \cos\left({k_0 \Delta n z} \right) \bm \sigma_2 + \sin\left({k_0 \Delta n z} \right) \bm \sigma_1 \right] \cdot \bm{u} \\ - \left(\frac{\partial\theta}{\partial z}\right)^2   \bm{u}=0,   
  \label{eq:rotated_only_Pauli_matrices}
\end{multline}
where we introduced $F(z)= \overline{n}  \cos\left({k_0 \Delta n z} \right) + i\frac{\Delta n}{2}  \sin\left({k_0 \Delta n z} \right)$ and $G(z)= \overline{n}  \sin\left({k_0 \Delta n z} \right) - i\frac{\Delta n}{2}  \cos\left({k_0 \Delta n z} \right)$. 
We are interested in the resonant case when the external modulation given by $\theta(z)$ is synchronized with the natural oscillation of the polarization setting $\Lambda=\lambda/\Delta n$. From Eq.~\eqref{eq:rotated_only_Pauli_matrices} this corresponds to a continuous-wave component coming from the terms depending on $\theta(z)$. The inversion operator of the matrix factor in front of $\partial_z \bm u$ is
\begin{equation} \label{eq:inversion_operator}
  \left(\overline{n} \bm I -\frac{\Delta n}{2} \bm \sigma_3 \right)^{-1} = \frac{1}{\overline{n}} \frac{1}{1-\left(\frac{\Delta n}{2\overline{n}} \right)^2} \left( \bm I + \frac{\Delta n }{2 \overline{n}} \bm \sigma_3 \right) .  
\end{equation}
Applying the inversion operator \eqref{eq:inversion_operator} to \eqref{eq:rotated_only_Pauli_matrices} we find
\begin{multline}
   2ik_0 \overline{n} \frac{\partial \bm{u}}{\partial z} + \\   K \left( \bm I + \frac{\Delta n }{2 \overline{n}} \bm \sigma_3 \right) \left[  X \bm \sigma_2 + Y \bm \sigma_1 - \left(\frac{\partial\theta}{\partial z}\right)^2 \right] \cdot   \bm{u}=0,    
  \label{eq:rotated_only_Pauli_matrices_explicit}
\end{multline}
where we set
\begin{align}
    X &= 2  k_0\frac{\partial\theta}{\partial z} F(z) - i\frac{\partial^2\theta}{\partial z^2} \cos\left({k_0 \Delta n z} \right) ,  \label{eq:definition_X}\\
    Y &=  2  k_0\frac{\partial\theta}{\partial z} G(z) - i \frac{\partial^2\theta}{\partial z^2} \sin\left({k_0 \Delta n z} \right) , \label{eq:definition_Y}\\
    K&=\left[1-\left(\gamma/2 \right)^2\right]^{-1}. \label{eq:definition_K}
\end{align}
For the sake of compactness, we introduce the normalized anisotropy $\gamma= \Delta n /\overline{n}$.
Computation of the operator multiplication in Eq.~\eqref{eq:rotated_only_Pauli_matrices_explicit} yields
\begin{multline}
     i \frac{\partial \bm u }{\partial z}=K
     \frac{ i (\gamma/2) X - Y }{2k_0 \overline{n}} \bm \sigma_1 \cdot \bm u - K\frac{ X +  i (\gamma /2) Y }{2k_0 \overline{n}} \bm \sigma_2 \cdot \bm u \\
     +  \frac{K}{2k_0 \overline{n}} \left( \bm I + \frac{\gamma }{2} \bm \sigma_3 \right) \left(\frac{\partial\theta}{\partial z}\right)^2 \cdot   \bm{u}
     \label{eq:rotated_effective_magnetic}
\end{multline}
By expanding $K$ in a power series of the normalized anisotropy $\gamma$, Eq.~\eqref{eq:rotated_effective_magnetic} can be recast as a power series of $\gamma$ itself. Before doing that, it is convenient to introduce the phase retardation $\zeta = k_0\Delta n z$, i.e., to normalize the propagation distance with respect to the natural rotation of the polarization vector. Equations~\eqref{eq:definition_X}, \eqref{eq:definition_Y} and \eqref{eq:definition_K} can then be expressed as a quadratic polynomial in $\gamma$
\begin{align}
    \frac{X}{\overline{n}^2} &= k_0^2 \left[ a_X(\theta) \gamma + b_X(\theta) \gamma^2 \right],  \label{eq:definition_X_gamma}\\
   \frac{Y}{\overline{n}^2} &= k_0^2 \left[  a_Y(\theta) \gamma + b_Y(\theta) \gamma^2 \right] , \label{eq:definition_Y_gamma}\\
    K&\approx 1 + \left( \frac{\gamma}{2}\right)^2.
\end{align}
The new terms defined in the above equations are
\begin{align}
     a_X(\zeta,\theta)&= 2  \cos\zeta \frac{\partial \theta}{\partial \zeta},  \label{eq:aX}\\
     b_X(\zeta,\theta)&= i \left( \sin\zeta \frac{\partial \theta}{\partial \zeta} - \cos\zeta \frac{\partial^2 \theta}{\partial \zeta^2} \right), \label{eq:bX} \\
     a_Y(\zeta,\theta)&= 2  \sin\zeta \frac{\partial \theta}{\partial \zeta},  \label{eq:aY}\\
     b_Y(\zeta,\theta)&= -i \left( \cos\zeta \frac{\partial \theta}{\partial \zeta} + \sin\zeta \frac{\partial^2 \theta}{\partial \zeta^2} \right) \label{eq:bY}.
\end{align}
Next we expand Eq.~\eqref{eq:rotated_effective_magnetic} in a power series of $\gamma$, halting the series to the linear order. Equation~\eqref{eq:rotated_effective_magnetic} then yields 
\begin{multline}
     i  \frac{\partial \bm u }{\partial \zeta}=-
     \frac{1}{2} \left(a_Y \bm \sigma_1 +a_X  \bm \sigma_2  \right) \cdot \bm u \\
     + \frac{\gamma}{2} \left[ \left(\frac{i a_X}{2} - b_Y \right) \bm \sigma_1 - \left(\frac{i a_Y}{2} + b_X   \right)  \bm \sigma_2 + \left(\frac{\partial \theta}{\partial \zeta}\right)^2 \right] \cdot \bm u .
     \label{eq:rotated_effective_magnetic_first_order}
\end{multline}
Using Eqs.~\eqref{eq:aX} and \eqref{eq:aY}, for small values of the anisotropy Eq.~\eqref{eq:rotated_effective_magnetic_first_order} turns into
\begin{equation}
     i  \frac{\partial \bm u }{\partial \zeta}=-\frac{\partial \theta}{\partial \zeta}
      \left[\sin\left(\zeta\right) \bm \sigma_1 +\cos\left(\zeta\right)  \bm \sigma_2  \right] \cdot \bm u.
\end{equation}
We now consider the resonant case, where the optic axis is periodically modulated with a period equal to $\lambda/\Delta n$ in the real space. Thus, after taking a sinusoidal oscillation in the form $\theta(\zeta)=\Gamma_0 \sin\left( \zeta \right)$, we find 
\begin{equation}
     i  \frac{\partial \bm u }{\partial \zeta}=-\frac{\Gamma_0}{2}
      \left[{\sin2\zeta}\ \bm \sigma_1 + \left(1+\cos2\zeta\right)  \bm \sigma_2  \right] \cdot \bm u   .
     \label{eq:rotated_effective_linear_order}
\end{equation}
From the Bloch-Floquet theorem, the quasi-mode can be expressed as
\begin{equation}
   \bm  u(\zeta)= e^{i\beta \zeta} \sum_m {\bm u_m e}^{i m\zeta}, \label{eq:ansatz_bloch}
\end{equation}
where $\beta$ is the associated eigenvalue.
Next step is inserting the ansatz Eq.~\eqref{eq:ansatz_bloch} into Eq.~\eqref{eq:rotated_effective_linear_order}. For each integer $m$ the following relation is found out
\begin{multline}
\beta \bm u_m = -m \bm u_m +
    \frac{\Gamma_0}{4}[2\bm \sigma_2 
    \bm u_m + \left(\bm \sigma_2 - i\bm \sigma_1 \right) \bm u_{m-2} \\ + \left(\bm \sigma_2 + i\bm \sigma_1 \right) \bm u_{m+2}  ]. \label{eq:eigenproblem_bloch}
\end{multline}
Equation~\eqref{eq:eigenproblem_bloch} shows that the components $u_m$ of different parity (i.e., the terms $u_m$ corresponding to $m$ either even or odd) form two independent sets of values.
In the case of small rotations (i.e., small $\Gamma_0$), the oscillations of the field amplitude are small, that means, $\bm u_m\approx 0$ for $m\neq 0$. Eq.~\eqref{eq:eigenproblem_bloch} then provides
\begin{equation}
    \beta \bm u_0 = \frac{\Gamma_0}{2} \bm \sigma_2 \cdot \bm u_0. \label{eq:eigenvalue_u0}
\end{equation}
From the latter it is straightforward to find that the eigenvectors are the two CPs $\left|L\right>$ and $\left|R\right>$, with the associated eigenvalues $\beta=\pm \Gamma_0/2$. In the real space coordinates, the phase delay acquired at each HWP length is $\pi \Gamma_0$, in agreement with the numerical simulations shown in Fig.~\ref{fig:plane_waves_eigenvectors}. \\
At the next order, we have $u_{\pm 1}\neq 0$. In this case Eq.~\eqref{eq:eigenproblem_bloch} provides an additional eigenvalue equation
\begin{equation}
\Delta \beta \ \bm u_{\pm 1} =  
    \left[\frac{\Gamma_0}{2}\bm \sigma_2 -
      \left(\beta_0 \pm 1\right) \bm I \right] \bm u_{\pm 1}, \label{eq:eigenproblem_u1}
\end{equation}
where we supposed $\beta\approx \beta_0 + \Delta \beta$, and where $\beta_0=\Gamma_0/2$ is the eigenvalue at the lowest approximation order, as determined by Eq.~\eqref{eq:eigenvalue_u0}. The eigenvectors are still circularly polarized. The two eigenvalues are $\pm \left(\Gamma_0 +1\right)$ and $\pm 1$, respectively, the sign depending on the sign of $m$. The solutions $\left|\Delta \beta\right|=\Gamma_0+1$ are not acceptable because they are inconsistent with the full eigenvalue equation~ \eqref{eq:eigenproblem_bloch}. On the other side, solutions featuring $|\Delta \beta|=1$ are acceptable because they imply a shift of $2\pi$ in the eigenvalue according to Eq.~\eqref{eq:ansatz_bloch}, thus representing the same solution according to the ansatz expressed by Eq.~\eqref{eq:eigenvalue_u0}. Summarizing, the simultaneous solution of Eq.~\eqref{eq:eigenvalue_u0} and Eq.~\eqref{eq:eigenproblem_u1} tells us that the beam in $z=0$ is circularly polarized, with an eigenvalue equal to $\beta_0$. An additional component with $|m|=1$ is present, providing a small change in the beam polarization, even in the rotated framework. This latter oscillation is actually responsible for the non-vanishing $S_1$ even in $z=0$, see Fig.~\ref{fig:plane_waves_eigenvectors_evolution}(a). To conclude, we notice that this simplified approach does not allow to quantify the relative weight of the two components $\bm u_0$ and $\bm u_{\pm 1}$, the latter evidently requiring the components $\bm u_m$ for $|m|>1$ to be accounted for. \\
\begin{figure}
    \centering
    \includegraphics[width=\linewidth]{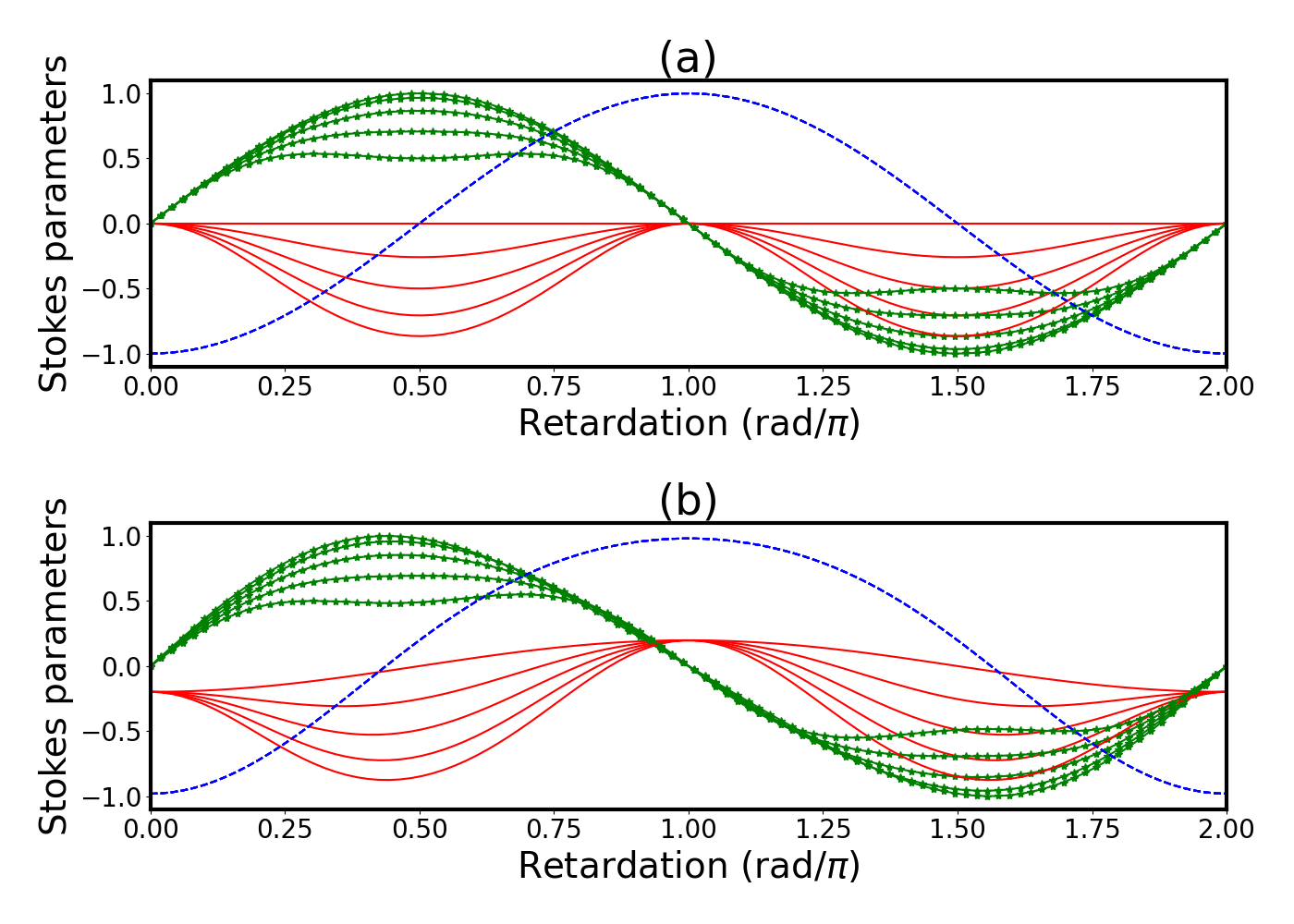}
    \caption{(a) The Stokes parameters versus the phase retardation for a CP wave in the rotated framework rotated back to the laboratory frame. Only $\bm u_0$ is non-vanishing in this case. (b) Stokes parameters versus phase retardation when $\bm u_1=-0.1\bm u_0$. In both panels the blue dashed curve is $S_3$, $S_2$ the green lines with symbol, $S_1 $ the solid red lines. The values used for $\Gamma$ are $0^\circ$, 7.5$^\circ$, $15^\circ$, $22.5^\circ$, $30^\circ$.}
    \label{fig:stokes_after_rotation}
\end{figure}
Figure~\ref{fig:stokes_after_rotation} compares the Stokes parameters in the laboratory framework when only $\bm u_0$ is non-vanishing [panel (a)] to the case where a $\bm u_1$ component with a $10\%$ amplitude of $\bm u_0$ is present [panel (b)]. Comparison with the full numerical simulations is discussed in the main text in Sec.~\ref{sec:quasimodes_1D}. To help the comparison, here we stress out the computation to connect the rotated and the laboratory framework. The wavefunction in the laboratory framework is linked to the coefficients $\bm u_m $ via
\begin{align}
    \bm \psi & = e^{i\beta\zeta} e^{ik_0 n_\bot \zeta} \times \nonumber \\  &\left(\begin{array}{c} 
        \cos\theta \ \sum_m u_{x,m}e^{im\zeta} -  \sin\theta\ \sum_m u_{y,m}e^{i(m+1)\zeta} \\ 
        \sin\theta \ \sum_m u_{x,m}e^{im\zeta} +  \cos\theta \ \sum_m u_{y,m}e^{i(m+1)\zeta}
        \end{array} \right). \label{eq:rotation_back}
\end{align}
In the limit of small angle $\theta=\Gamma(\xi) \sin\zeta$, considering only $\bm u_0\neq 0$ we find
\begin{equation}
    \bm \psi_0 \approx \left(\begin{array}{c} 
         u_{x,0} \\ 
           u_{y,0} e^{i\zeta}
        \end{array} \right) +\frac{\Gamma}{2i}
        \left(\begin{array}{c} 
           - u_{y,0}e^{i\zeta}\\
           u_{x,0}
        \end{array}\right), 
\end{equation}
where the common phase $(\beta+k_0 n_\bot)\zeta$ has been removed for the sake of clarity. For $\Gamma=0$, the limit of a homogeneous wave plate is correctly retrieved. When considering the terms for $|m|=1$ we obtain
\begin{align}
     \bm \psi_0 \approx \ldots + &\left(\begin{array}{c} 
         u_{x,1} e^{i\zeta} + u_{x,-1} e^{-i\zeta} \\ 
            u_{y,1} e^{2i\zeta} + u_{y,-1} 
        \end{array} \right) + \nonumber \\\frac{\Gamma}{2i}
        & \left(\begin{array}{c} 
           - u_{y,-1} - u_{y,1} e^{2i\zeta}\\
           u_{x,-1} e^{-i\zeta} + u_{x,1} e^{i\zeta}
        \end{array}\right).  
\end{align}
The presence of terms not explicitly dependent on $\zeta$ demonstrates how the harmonics for $|m=1|$ affect the average value of $\bm \psi_0$, in agreement with Fig.~\ref{fig:stokes_after_rotation}(b).


\section{Jones matrix in a layered twisted material}
\label{sec:Jones_formalism}
In the circular basis $(\left|L\right>,\left|R\right>)$ and for unidirectional light propagation, the Jones matrix for a transversely homogeneous slab of uniaxial material of thickness $\delta$ and twisted by an angle $\theta$ is 
\begin{align}
        \bm J(\delta,\theta) = & \nonumber \\  e^{i\overline{n}k_0 \delta} & \left(\begin{array}{cc}
        \cos\left(\frac{k_0 \Delta n \delta}{2}\right) &
        -i\sin\left( \frac{k_0 \Delta n \delta}{2} \right) e^{2i\theta} \\ -i\sin\left( \frac{k_0 \Delta n \delta}{2}\right) e^{-2i\theta} & \cos\left(\frac{k_0 \Delta n \delta}{2}\right) 
    \end{array} \right) = \nonumber \\
    e^{i\overline{n}k_0 \delta} & e^{-i\frac{k_0 \Delta n \delta}{2} \left( \hat{s} \cdot \bm \sigma  \right)} = e^{i\overline{n}k_0 \delta \left[1-\frac{\Delta n}{2\overline{n}}  \left( \hat{s} \cdot \bm \sigma \right)  \right]} \label{eq:Jones_LR}
\end{align}
where $\hat{s}(\theta)=\cos\left(2\theta \right) \hat{x} - \sin(2\theta) \hat{y}$. When the eigenvalues of the exponential matrix are computed, we correctly retrieve the ordinary and extraordinary plane waves as eigensolution of the system, but rotated by an angle $\theta$ with respect to the framework $xy$. 

For a stack of infinitely thick layers of overall thickness $L$, the total transfer function is given by the multiplication of $N$ matrices $\bm J(\delta,\theta_m)$, each of them calculated in the limit  $\delta\rightarrow 0$. This approach is valid in the limit of slow variations for the angle $\theta$ on the scale $\lambda/\Delta n$, i.e., in the adiabatic limit. Dubbing $L$ the overall length of the twisted material, we have $\delta=L/N$; finally, in the limit of infinitely-thin layers the transmission matrix in the absence of back-reflections is
\begin{align}
 \bm J_{total} = &   \nonumber \\ \lim_{N\rightarrow \infty}  {\prod_{m=1}^{N}} &\left[ \bm I + i k_0 \overline{n}\frac{L}{N}   \left(\begin{array}{cc}
        1 & -\frac{  \gamma  }{2} e^{2i\theta_m}   \\
        -\frac{  \gamma  }{2} e^{-2i\theta_m} & 1  
    \end{array} \right) \right]  .  \label{eq:Jones_LR_overall}
\end{align}
From Eq.~\eqref{eq:Jones_LR_overall} we deduce that the propagation of a plane wave in a longitudinally-twisted geometry can be normalized with respect to the normalized anisotropy $\gamma=\Delta n/\overline{n}$. \\ 
The fundamental properties of the solutions to Eq.~\eqref{eq:Jones_LR_overall} can be better visualized if we use the last expression in Eq.~\eqref{eq:Jones_LR}. We find that
\begin{equation}
 \bm J_{total} =  e^{i\overline{n}k_0 L} \lim_{N\rightarrow \infty}  {\prod_{m=1}^{N}} e^{-i\frac{k_0 \Delta n \left[ \hat{s}(\theta_m) \cdot \bm \sigma \right] }{2}\frac{L}{N} }  .  \label{eq:Jones_integrated}
\end{equation}
According to Eq.~\eqref{eq:Jones_integrated}, if $\theta=\theta(\zeta)$ the optical propagation depends only on the phase retardation $k_0 \Delta n \delta$, except for a phase term corresponding to the the dynamic phase of a CP wave. Once $\bm J_{total}$ is known, the corresponding eigenmodes (i.e., the polarization at each FWP -Full Wave Plate- distance) and the eigenvalues (i.e., the associated geometric phase) can be numerically computed using standard algebraic methods. 

\section{Modelling of the transverse coupling}
\label{sec:transverse_coupling}

We start by considering only the right hand side (RHS) of Eq.~\eqref{eq:maxwell_rotated_inho}; let us call it the operator $\hat{L}$. Applying the SVEA (see the definition of $\bm u$ before Eq.~\eqref{eq:rotated_SVEA_PW_2} in Appendix~\ref{sec:plane_wave_model}) we find
\begin{equation}
    \hat{L}= -   \frac{\partial^2 \bm{u}}{\partial x^2} + \left(\frac{\partial\theta}{\partial x} \right)^2 \bm{u}   + i \frac{\partial^2 \theta}{\partial x^2}  \bm{\tilde{\sigma}}_2\cdot \bm{u} + 2i  \frac{\partial\theta}{\partial x} \bm{\tilde{\sigma}}_2 \cdot \frac{\partial \bm{u}}{\partial  x}.
\end{equation}
The aim of the current section is to develop the transverse coupling alone, considering the minimal coupling with the evolution of the field along $z$.
From Eq.~\eqref{eq:rotated_only_Pauli_matrices} we can use the simplified equation 
\begin{equation}
2ik_0 \left(\overline{n} \bm I -\frac{\Delta n}{2} \bm \sigma_3 \right) \cdot \frac{\partial \bm{u}}{\partial z}= \hat{L},
\end{equation}
i.e., we account only for the term providing the first derivative of the field along the propagation coordinate $z$. Application of the operator defined by Eq.~\eqref{eq:inversion_operator} to both sides provides
\begin{multline}
   \frac{2ik_0  \overline{n}}{K} \frac{\partial \bm{u}}{\partial z} =
   \left( \bm I +\frac{\gamma}{2} \bm \sigma_3 \right) \cdot \hat{Q}(x) \bm u     + \\ 
    \left[\left(\bm \sigma_2 -  \frac{i\gamma}{2}\bm \sigma_1 \right) \cos\left({k_0 \Delta n z} \right) + \left(\bm \sigma_1 +  \frac{i\gamma}{2}\bm \sigma_2 \right) \sin\left({k_0 \Delta n z} \right) \right] \cdot \\
    \hat{P}(x) \bm u,
  \label{eq:rotated_diffraction}
\end{multline}
where $K$ has been defined in Eq.~\eqref{eq:definition_K} and
\begin{align}
    \hat{Q}(x) &=  -   \frac{\partial^2 }{\partial x^2} + \left(\frac{\partial\theta}{\partial x} \right)^2    \\
    \hat{P}(x)  &= i\frac{\partial^2 \theta}{\partial x^2}  +2i\frac{\partial \theta}{\partial x} \frac{\partial }{\partial x}
\end{align}
Introducing the normalized transverse coordinate $\eta=x/\lambda$ and the retardation $\zeta=k_0\Delta n z$, Eq.~\eqref{eq:rotated_diffraction} can be recast as
\begin{equation}
   i\gamma  \frac{\partial \bm{u}}{\partial \zeta} = \frac{1}{8\pi^2 \overline{n}^2} \left( \bm T_0 + \bm T_1 \gamma + \bm T_2 \gamma^2 + \ldots \right) \cdot \bm u.
  \label{eq:rotated_diffraction_normalized}
\end{equation}
Until the order $\gamma^2$ we find
\begin{align}
 \bm{T}_0 &=    \hat{Q}(\eta) \bm u + \left[ \bm \sigma_2 \cos\left( \zeta\right)+ \bm \sigma_1 \sin\left( \zeta\right)\right] \hat{P}(\eta) \bm u, \\
 \bm{T}_1 & = \frac{1}{2} \bm \sigma_3 \cdot \hat{Q}(\eta)  \bm u \nonumber\\ &\  +\frac{i}{2} \left[ \bm \sigma_2 \sin(\zeta) - \bm \sigma_1 \cos(\zeta) \right]\cdot \hat{P}(\eta) \bm u, \\
 \bm{T}_2 & = \left( \frac{1}{2}\right)^2 \bm T_0.
\end{align}
Equation~\eqref{eq:rotated_diffraction_normalized} explicitly states that the effects of diffraction can be described as a power expansion in the normalized anisotropy $\gamma$. We finally take the resonant case setting $\theta(\eta,\zeta)=\Gamma(\eta)\sin(\zeta)$. At the lowest order in $\gamma$ and considering only the averaged term along the propagation coordinate $\zeta$, the field evolves according to
\begin{multline}
 i\gamma  \frac{\partial \bm{u}}{\partial \zeta} \approx \frac{1}{8\pi^2 \overline{n}^2} \Bigg[  - \frac{\partial^2 \bm u}{\partial \eta^2} + \frac{1}{2}\left(\frac{\partial \Gamma}{\partial \eta} \right)^2 \bm u \\ + \frac{i \bm \sigma_1}{2} \cdot \left(\frac{\partial^2 \Gamma}{\partial \eta^2} \bm u + 2 \frac{\partial \Gamma}{\partial \eta} \frac{\partial \bm u}{\partial \eta} \right) \Bigg].
\end{multline}
The term proportional to $\bm \sigma_1$ can be eliminated by employing the gauge transformation $\bm u=e^{i\bm \sigma_1 \Gamma/2}\cdot \bm v$, in turn providing the final result
\begin{equation}
 i\gamma  \frac{\partial \bm{v}}{\partial \zeta} \approx \frac{1}{8\pi^2 \overline{n}^2} \left[  - \frac{\partial^2 \bm v}{\partial \eta^2} + \frac{1}{4}\left(\frac{\partial \Gamma}{\partial \eta} \right)^2 \bm v  \right].
 \label{eq:diffraction_after_gauge}
\end{equation}

\section{Derivation of the complete model}

We can now derive the whole model for the optical propagation combining the results derived in Appendix~\ref{sec:plane_wave_model} and in Appendix~\ref{sec:transverse_coupling}. Joining Eq.~\eqref{eq:rotated_diffraction_normalized} and Eq.~\eqref{eq:rotated_effective_magnetic_first_order}, in the normalized coordinate system $\xi\zeta$ we find
\begin{multline}
   i  \frac{\partial \bm{u}}{\partial \zeta} = -\frac{\partial \theta}{\partial \zeta}
      \left[{\sin\left(\zeta\right)}\ \bm \sigma_1 + \cos\left(\zeta\right)  \bm \sigma_2  \right] \cdot \bm u  \\ + \frac{1}{8\pi^2  \overline{n}^2 \gamma} \left( \bm T_0 + \bm T_1 \gamma + \bm T_2 \gamma^2 + \ldots \right) \cdot \bm u.
  \label{eq:complete_model}
\end{multline}
In the limit of small anisotropy and in the resonant case $H(\zeta)=\sin\left( \zeta \right)$, Eq.~\eqref{eq:complete_model} provides
\begin{multline}
   i \gamma  \frac{\partial \bm{u}}{\partial \zeta} =  -\frac{\gamma \Gamma(\eta)}{2}
      \left\{{\sin\left(2\zeta\right)}\ \bm \sigma_1 + \left[1+\cos\left(2\zeta\right) \right]  \bm \sigma_2  \right\} \cdot \bm u  \\ + \frac{1}{8\pi^2  \overline{n}^2} \Bigg\{
      -\frac{\partial^2 \bm u}{\partial \eta^2} + \frac{1}{2} \left( \frac{\partial \Gamma}{\partial \eta}\right)^2 \bm u + 
      \frac{i}{2} \Big\{\sin\left(2\zeta\right)\ \bm \sigma_2 + \\ \left[1+\cos\left(2\zeta\right) \right]  \bm \sigma_1  \Big\} \left(\frac{\partial^2 \Gamma}{\partial \eta^2}  +2\frac{\partial \Gamma}{\partial \eta} \frac{\partial }{\partial \eta} \right)\bm u  \Bigg\} .
  \label{eq:complete_model_small_gamma}
\end{multline}
Eq.~\eqref{eq:complete_model_small_gamma} describes the evolution of waves, including the beam variations occurring inside any single birefringence length. Eq.~\eqref{eq:Pauli_model} in the main text is then derived by rewriting the field as a Bloch wave and considering only the CW component. The last step is carried out in a simplified manner by averaging the $\zeta$-dependent coefficients over a birefringence length. For a more accurate approach, see Eq.~\eqref{eq:eigenproblem_bloch}. Finally, the terms explictly dependent on $i$ can be factored out by using a gauge transformation, in full analogy with what has been done to achieve Eq.~\eqref{eq:diffraction_after_gauge}.

\section{Details of the FDTD implementation}
\label{sec:appendix_FDTD}
The FDTD is run using a continuous source with a wavelength of $1~\mu$m. The switching parameters of the source are chosen such that to achieve the stationary solutions inside the temporal duration of our simulations.
To inject the quasi-mode as input on the FDTD simulations, we first generate a fictitious isotropic material with a refractive index profile matching the potential given by Eq.~\eqref{eq:potential_first_order}. The polarization is then transformed into circular by inserting a homogeneous layer of anisotropic material with thickness corresponding to a QWP. The dielectric permittivities of the QWP are taken identical to the twisted material to minimize the reflection at the input interface, the latter implying a change in the polarization actually transmitted into the structured material. \\
The time-average intensity is derived from the fields oscillating in time by either applying a low-pass Savitzky-Golay filter or by time averaging the electric field saved in one temporal oscillation (21 points are saved in one oscillation) after the stationary regime is achieved. We verified that the two approaches yield the same results, with the first method presenting some small residual oscillation along the propagation direction. Analogously, the Stokes parameters are retrieved by deriving the complex amplitude of the field from the positions of the maxima in the temporal oscillation of the field. Notice that this procedure tacitly assumes a negligible amount of back-reflection in the twisted material. Also in this case, a more robust procedure based upon best-fitting of the whole wavefunction along one period provides no substantial differences.

\begin{figure}
    \centering
    \includegraphics[width=\linewidth]{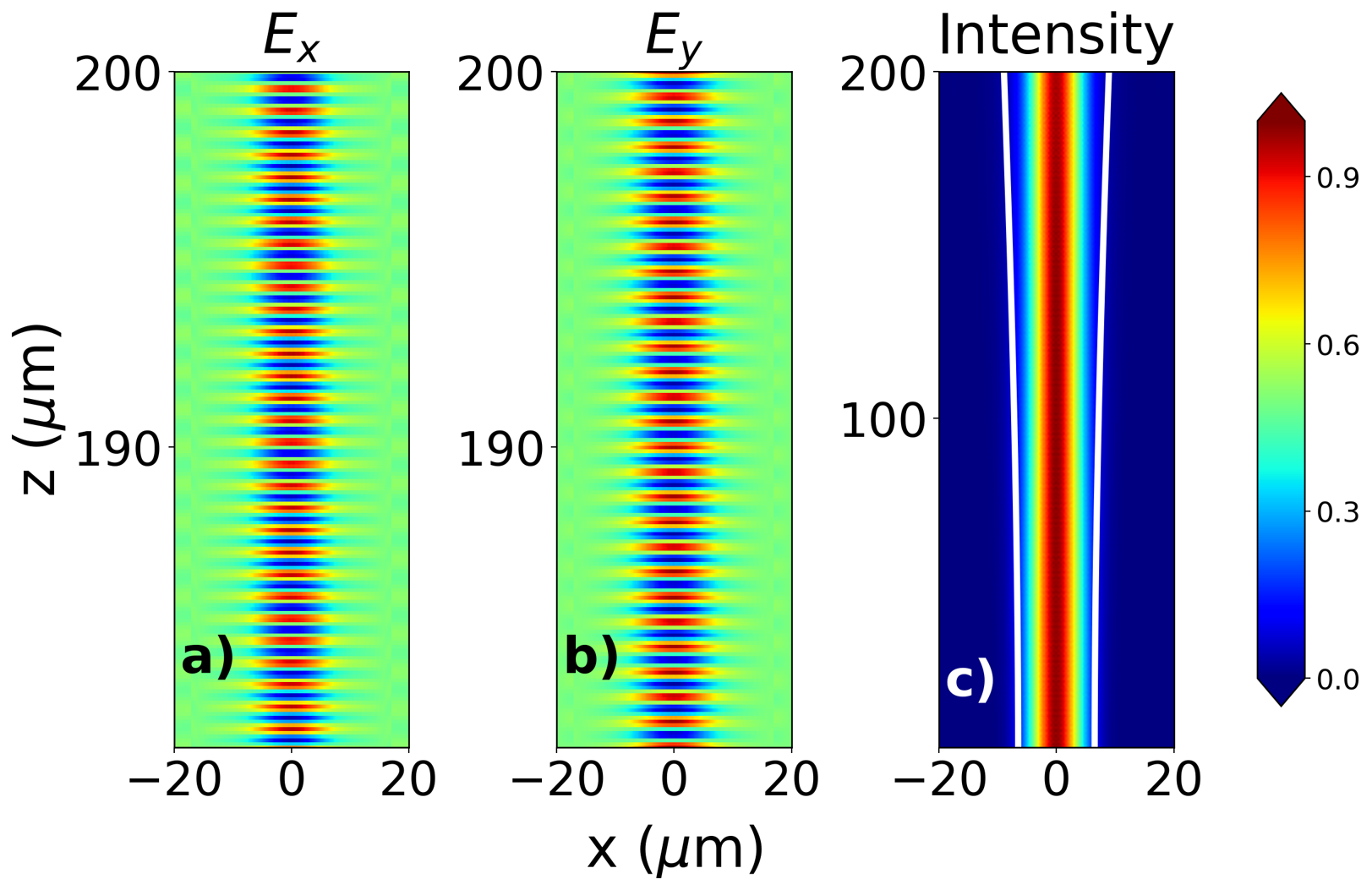}
    \caption{Long FEM simulations for small rotation angles. Maps on the plane $xz$ of the electric field components (a) $E_x$,  (b) $E_y$, and (c) of the corresponding time-averaged Poynting vector component along the propagation distance $z$. A snap-shot of the time-dependent electric fields for $180~\mu$m$<z<200~\mu$m is shown in panels (a,b). The maximum rotation angle is $\Gamma_0=1^\circ$ and $w_D=8~\mu$m. Input is a Gaussian of width $6.6~\mu$m.}
    \label{fig:FEM_long}
\end{figure}

\begin{figure}
    \centering
    \includegraphics[width=\linewidth]{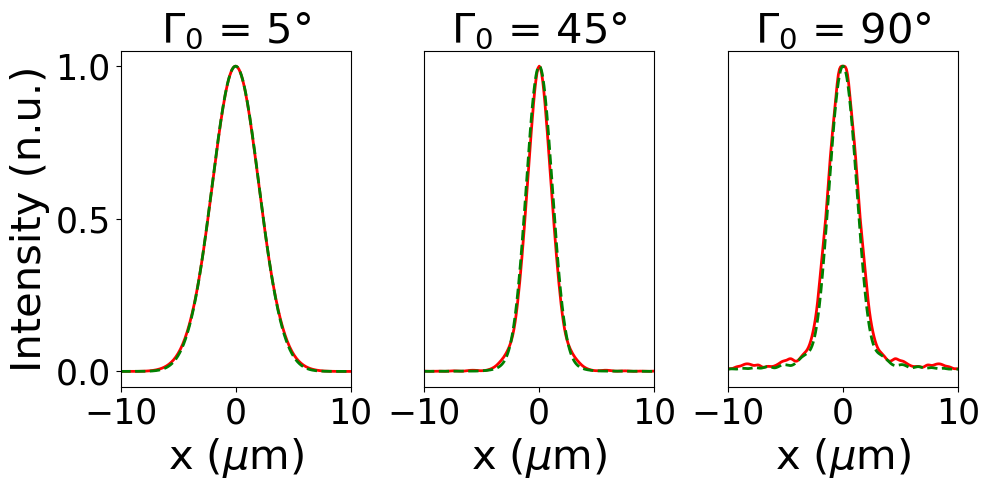}
    \caption{Comparison between FDTD and FEM simulations. The normalized intensity cross-section in $z=50~\mu$m is plotted versus $x$ for three different values of $\Gamma_0$, the latter being labelled at the top of each panel; the width of the twisting distribution is $w_D=8~\mu$m. Solid red and green dashed lines correspond to FDTD and FEM, respectively. In both the simulators, the input is a circularly polarized Gaussian beam with a width equal to the effective fundamental mode.}
    \label{fig:FDTD_VS_FEM_profiles}
\end{figure}

\section{FEM simulations and comparison with FDTD results}
\label{sec:appendix_FEM}

\begin{figure}
    \centering
    \includegraphics[width=\linewidth]{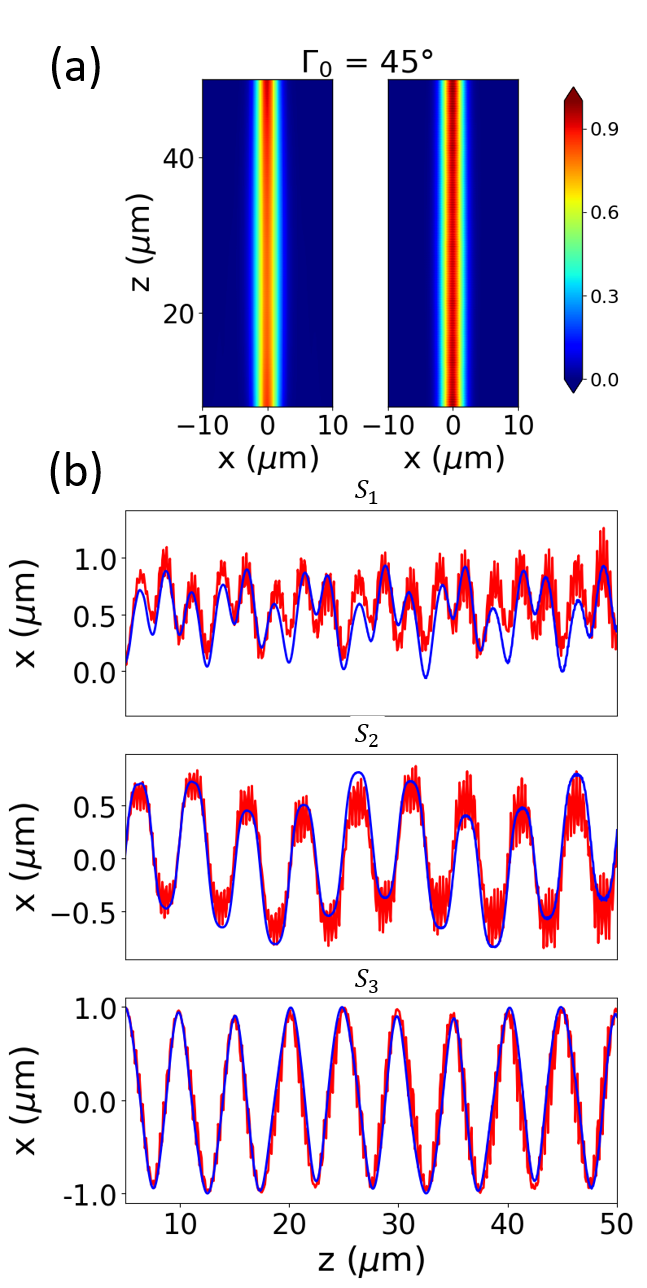}
    \caption{Comparison between FDTD and FEM simulations. (a) Intensity distribution for a Gaussian input of waist $2.4~\mu$m over the plane $x$ for $\Gamma_0=45^\circ$ and $w_D=8~\mu$m, computed with FEM (left side) and FDTD (right side). (b) Corresponding evolution versus $z$ of the Stokes parameters on the beam axis $x=0~\mu$m; blue and red lines correspond to FDTD and FEM simulations, respectively.}
    \label{fig:FDTD_VS_FEM}
\end{figure}

During our numerical efforts we found out that FDTD simulations for very small angles (lower than $5^\circ$) do not converge properly, even with spatial steps of about 20~nm. In particular, the intensity profile and the two Stokes parameters $S_2$ and $S_3$ achieve a convergence, but the simulations predict a spurious $S_1$ component encompassing a non-vanishing error versus the coarseness of the numerical grid is present. Curiously, such a behavior does not take place for large angles. After several tests, we deduced that the error comes from the interpolation function used by the program to interpolate the given point-dependent function for the dielectric tensor, with the most critical point being the interface between the twisted material and the QWP layer. To verify the accuracy of the numerical results for small $\Gamma_0$, we used COMSOL Multiphysics\textsuperscript{\textregistered} to simulate the light propagation, but using a Gaussian beam at the input, with a waist equal to the theoretical value predicted from Eq.~\eqref{eq:potential_first_order}. In COMSOL we employed the frequency domain calculation available in the Wave Optics module. We first simulated the case $\Gamma_0=1^\circ$ and $w_D=8~\mu$m over a long cell (length $200~\mu$m), see Fig.~\ref{fig:FEM_long}. The confinement occurs as shown in Fig.~\ref{fig:FDTD_survey}(a), with the Stokes parameters converging in a smooth way. To save time, we then switched to shorter cells (length $60~\mu$m along the propagation direction) to validate the FDTD simulations versus the maximum rotation angle $\Gamma_0$. To further relax the numerical requirements, we focused on the case $w_D=8~\mu$m. In both the simulators, we took a Gaussian beam placed in $z=0~\mu$m in air, whereas the twisted material starts at $z=2~\mu$m. Figure~\ref{fig:FDTD_VS_FEM_profiles} shows the intensity cross-section computed in $z=50~\mu$m with FDTD (green dashed lines) and FEM (red solid lines). 
A very good agreement is found between the two methods. Small differences can be seen on the tails, with the FEM case showing some ripples. This is due to the PML (Perfectly Matched Layer) boundary conditions, inducing non-negligible back reflections from the edges of the grid. Such reflections increases with $\Gamma_0$, explaining the growing differences in the tails of the predicted field. Figure~\ref{fig:FDTD_VS_FEM} provides more details. The full intensity distribution in the plane $xz$ shows some small difference in the beam amplitude, see Fig.~\ref{fig:FDTD_VS_FEM}(a). Beyond the numerical reflections discussed above, small discrepancies can be ascribed to slightly different definitions of the input Gaussian beam. The Stokes parameters versus $z$ are very smooth in the case of the FDTD, whereas fast variations are observed in the FEM results, see Fig.~\ref{fig:FDTD_VS_FEM}(b). This validates our previous statement that in the FEM simulations the spurious numerical reflection from the grid edges are much stronger than in the FDTD, at least for the PML parameters (default setting) we chose. Indeed, the back reflections are greatly reduced when an air buffer is inserted between the PML and the twisted material (condition we used in the plotted results), demonstrating that the standard PML does not work properly in our case.  
Summarizing, the case of light propagating in a twisted anisotropic material is highly demanding from a numerical point of view, even in the linear regime: extreme attention should be paid when numerical simulations are performed in these geometries.  


\bibliography{references}

\end{document}